\documentclass[12pt,tightenlines,aps,prd,nofootinbib,superscriptaddress,eqsecnum]{revtex4}

\usepackage{amsmath}
\usepackage{amssymb}
\usepackage{bm}
\usepackage{graphicx}
\usepackage{xcolor}
\usepackage{orcidlink}
\usepackage{algorithm}
\usepackage{algpseudocode}

\newcommand{\Tr}{\operatorname{Tr}}
\newcommand{\ReTr}{\operatorname{Re\,Tr}}

\newcommand{\muhat}{\hat{\mu}}
\newcommand{\nuhat}{\hat{\nu}}

\newcommand{\SU}{\mathrm{SU}}
\newcommand{\su}{\mathfrak{su}}

\begin{document}

\title{Sampling $\SU(N)$ gauge theory on a 2D lattice from independent plaquettes via holonomies and corner reweighting}
\author{Javad Komijani\,\orcidlink{0000-0002-6943-8735}}
\date{\today}

\begin{abstract}
In the Wilson formulation of lattice gauge theory, the fundamental degrees of freedom are group-valued link variables, while the action is a sum over the trace of the plaquettes, the smallest Wilson loops.
For generative sampling methods such as normalizing flows, this poses a challenge: the distribution of an individual plaquette is easy to model, but mapping sampled plaquettes to the links is the obstruction.
We explore a statistical way around this in two dimensions for the $\SU(N)$ gauge theory with the Wilson plaquette action.
The action can be written in terms of holonomy variables, which determine the links once a consistency condition at the four corners of an extended lattice is satisfied.
Using a boundary condition that only affects the holonomies, we trade this consistency condition for a conditional sampling problem, which reduces to sampling $X,Y\in\SU(N)$ given the group commutator $Z=XYX^\dagger Y^\dagger$, where $Z\in\SU(N)$ depends on the four corners.
Plaquettes are sampled independently, and the resulting configurations carry weights due to the additional conditional sampling.
We obtain the density of $Z$, which determines these weights, in closed form for $\SU(2)$ and $\SU(3)$.
A normalizing flow models the single-plaquette distribution for $\SU(2)$ and $\SU(3)$;
for the group-commutator problem we use a closed-form sampler for $\SU(2)$ and a trained normalizing flow for $\SU(3)$.
In our tests on $2\times2$ and $32\times32$ lattices at two couplings each, per-plaquette acceptance rates exceed $99\%$ and the effective sample size of the final weights is moderate, typically above one half.
\end{abstract}

\maketitle

\section{Introduction}
\label{sec:intro}

Lattice gauge theory, introduced by Wilson~\cite{Wilson:1974sk}, is the
standard non-perturbative framework for simulating gauge theories, including the Yang--Mills action. The fundamental degrees of freedom are
link variables, and the simplest discretization, i.e. the Wilson plaquette action, sums over the
smallest closed loops on the lattice.
In two dimensions, Yang--Mills theory is exactly solvable~\cite{Migdal:1975zg,Drouffe:1983fv,Rusakov:1990rs,Witten:1991we,Griguolo:2022vdx}.
A standard lattice derivation is based on a group-character expansion of the single-plaquette Boltzmann factor, with link integrations enforcing equality of the representation labels of neighboring plaquettes.

The solution, however, does not directly provide the
configuration-level map needed for generative sampling.
A natural question is whether an exact trivializing
map~\cite{Luscher:2009eq}
can be constructed for two-dimensional Yang–Mills theory.
There have been several papers recently tackling the sampling problem
using generative modeling, such as the methods of normalizing flows
and diffusion models~\cite{Kanwar:2020xzo, Boyda:2020hsi, Favoni:2020reg, Bacchio:2022vje, Komijani:2025yjz, Aarts:2026zzr, Alharazin:2026lcb, Komijani:2026lan, Komijani:2026mlx}.
In particular, for flow-based generative methods,
the distribution of a single plaquette matrix is easy to model,
but mapping sampled plaquettes to the underlying links is the
obstruction.
One can exclude one plaquette from sampling, compute all links,
and then weight the configuration based on the weight coming from the
remaining plaquette. This leads to an effective sample size that
depends drastically on the gauge coupling.

We circumvent the problem by giving a statistical
construction in two dimensions, a weighted map between plaquettes and links.
To this end, we employ an equivalent formulation of
lattice gauge theory using \emph{string} and \emph{corner} variables
\cite{Bars:1978ry,Batrouni:1984rb}; building on those
works, we introduce prelinks $V_\mu(x)$, defined on an extended
lattice, and the holonomy
\begin{equation}
  h_{\mu 0}(x)=V_\mu(x)V_0^\dagger(x).
\end{equation}
Unlike link variables, which are bilocal in the sense that their gauge
transformation depends on two neighboring sites, prelinks are local up
to an additional semi-global freedom, the same for all prelinks
along a line.
This semi-global freedom allows us to impose a boundary condition on
the prelinks that differs from the original boundary condition of the
links; this freedom is exploited in this paper.

The plaquette variables can be written in terms of the holonomy variables in
a unique way, but recovering the link field from a given
holonomy configuration requires resolving a consistency condition at
the four corners of the extended lattice; the same obstruction
Batrouni and Halpern \cite{Batrouni:1984rb} identified on the
periodic lattice but left as an imposed condition.

This condition is a hard, group-valued delta-function constraint, so
reweighting samples directly against it is impractical. In two dimensions
it is the only constraint on the holonomy variables. Using a twisted
boundary condition on the prelinks, without compromising the original
boundary condition on the links, we trade it for two $\SU(N)$ variables
$(X, Y)$ sampled conditional on
\begin{equation}
  Z = XYX^\dagger Y^\dagger,
\end{equation}
with $Z\in\SU(N)$ built from the corner holonomies.

We do this by \emph{encoding} $(X,Y)$ into three $\SU(N)$ matrices
$(D,\Omega,Z)$, with $D$ diagonal, and \emph{decoding} them back in closed
form for $\SU(2)$ and $\SU(3)$; this turns the corner condition from a
delta function into an ordinary probability density
$p_\text{encode}(D, \Omega, Z(h))$. The full prelink measure then
factorizes as
\begin{equation}
  p(V)\,dV\,dX\,dY = p(h)\,p_\text{encode}(D,\Omega, Z(h))\,dh\,dD\,d\Omega,
\end{equation}
with $p(h)$ the open-boundary holonomy theory, which can be fully
factorized and sampled independently, plaquette by plaquette.

This is not a map but a statistical construction: computing
$(V, X, Y)$ from $(h,D,\Omega)$ is deterministic, but $(D,\Omega)$ must
itself be drawn from $P(D,\Omega\mid Z(h))$ and the resulting samples
reweighted against it. For $\SU(2)$ we solve this conditional sampling
problem in closed form. For $\SU(3)$ this work is in progress, and for now we
set up the sampler with a normalizing flow. We also use normalizing flows for the single-plaquette
distribution.
We demonstrate the construction for $\SU(2)$ and $\SU(3)$,
and obtain a corner-correction effective sample size that is moderate, typically above one half, for $2\times2$ and $32\times32$ lattices at two couplings each.

Unlike the single-plaquette-exclusion baseline above, this effective sample size
does not depend on the gauge coupling directly, but only through the
dependence of $Z$ on it. With an exact corner sampler, which we have for $\SU(2)$, it is determined by
the density of $Z$ alone. This density depends on the lattice size and the
coupling, but approaches the Haar distribution for asymptotically large
lattices, where the effective sample size remains moderate.

The paper is organized as follows. In sec.~\ref{sec:params} we introduce
the link, prelink, and holonomy parametrizations. In
sec.~\ref{sec:constraint} we show how the corner condition leads to the
group-commutator problem for $(X,Y)$ and to the factorized measure. We
present the encoding of $(X,Y)$ and the conditional sampling of $(D,\Omega)$
given $Z$, and summarize the resulting sampling workflow. The tests are
presented in sec.~\ref{sec:simulations},
including comparing against Hybrid Monte Carlo (HMC)~\cite{Duane:1987de} simulations,
and we conclude in
sec.~\ref{sec:conclusion}.
\section{Links, Prelinks, and Holonomy}
\label{sec:params}

In this section we set up the link, prelink, and holonomy
parametrizations of the Wilson gauge action, similar to the
formulations introduced in refs.~\cite{Bars:1978ry,Batrouni:1984rb},
arriving at a corner constraint that is left unresolved in the current section and
addressed in sec.~\ref{sec:corner-resolution}.
We discuss the symmetries of these formulations in some detail, since one of them lets us trade the hard constraint for a softer sampling problem.

\subsection{Physical lattice and gauge links}
\label{sec:links}

Let $\Lambda=\{x: 0\le x_\mu<L_\mu\}$ be an $n$-dimensional hypercubic
lattice. We assume periodic boundary conditions throughout the
manuscript, but the choice of boundary condition is not essential
here. The fundamental
degrees of freedom are gauge links $U_\mu(x)\in\SU(N)$, representing
parallel transport from $x$ to $x+\muhat$. The Wilson gauge action is
\begin{align}
  S[U] &= -\frac{\beta}{N}\sum_{x\in\Lambda,\,\mu>\nu} \ReTr\, P_{\mu\nu}(x),
  \label{eq:wilson}
  \\
  P_{\mu\nu}(x) &= U_\mu(x)\,U_\nu(x+\muhat)\,U_\mu^\dagger(x+\nuhat)\,U_\nu^\dagger(x),
  \label{eq:plaq:U}
\end{align}
where $P_{\mu\nu}(x)$ denotes the plaquette, the smallest closed
Wilson loop.
The partition function is the Haar-measure integral over all links,
\begin{equation}
  \mathcal{Z} = \int \prod_{\mu}\prod_{x\in\Lambda} dU_\mu(x)\;e^{-S[U]},
  \label{eq:Z-links}
\end{equation}
invariant under the gauge transformation
\begin{equation}
  U_\mu(x) \to \Omega(x)\,U_\mu(x)\,\Omega^\dagger(x+\muhat),
  \qquad \Omega(x)\in\SU(N).
  \label{eq:link-gauge}
\end{equation}

\subsection{Minimally extended lattice and prelinks}
\label{sec:prelinks}

\begin{figure}[tbp]
  \centering
  \includegraphics[width=1\linewidth, trim=0cm 0cm 0cm 1cm, clip]{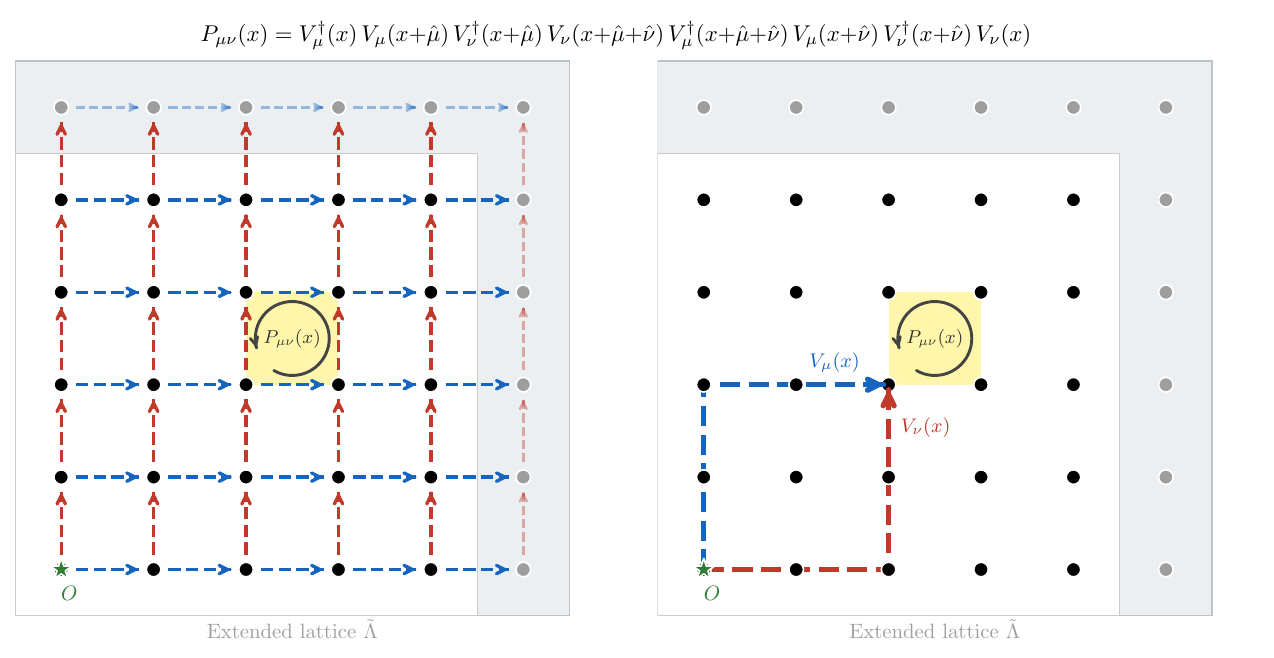}
  \caption{Prelinks $V_\mu(x)$ on the extended lattice $\tilde\Lambda$.
    The gray band shows the extension $\tilde\Lambda\setminus\Lambda$.
    Left: the straight-line recursion eq.~\eqref{eq:prelink-recursion},
    colored by direction ($\mu$ blue, $\nu$ red); the faded lines are
    the boundary copies of eq.~\eqref{eq:twisted-bc}, lying
    entirely inside the extension. Right: the L-shaped pattern arising
    from the scheme of eq.~\eqref{eq:Q-fix} for
    fixing the initial-value symmetry.}
  \label{fig:v_prelinks}
\end{figure}

The \emph{extended lattice} $\tilde\Lambda$ adds one extra site per
direction, $\tilde\Lambda=\{x: 0\le x_\mu\le L_\mu\}$. We also use
$\Lambda_{\perp,\mu}\subset\Lambda$ to denote the sites with $x_\mu=0$
in $\Lambda$, and introduce the \emph{minimal} extended lattice
\begin{equation}
  \tilde\Lambda_\text{minimal}
    = \Lambda\,\cup\,\left\{\bigcup_\mu\{x+L_\mu\muhat : x\in\Lambda_{\perp,\mu}\}\right\},
  \label{eq:minimal-extension}
\end{equation}
which adds, for each direction $\mu$ separately, only the single far
hyperplane. We insert into eq.~\eqref{eq:Z-links} the identity
\begin{equation}
  1 = \prod_{\mu} \prod_{x\in\Lambda_{\perp,\mu}}\int dV_\mu(x)
  \label{eq:trivial-insert}
\end{equation}
and define, recursively along each line in direction $\mu$,
\begin{equation}
  V_\mu(x+\muhat) = V_\mu(x)\,U_\mu(x), \qquad x_\mu=0,1,\dots,L_\mu-1,
  \label{eq:prelink-recursion}
\end{equation}
starting from the initial value just introduced. The link variable can
be recovered as
\begin{equation}
  U_\mu(x) = V_\mu^\dagger(x)\,V_\mu(x+\muhat), \qquad x\in\Lambda\,.
  \label{eq:U-from-V}
\end{equation}
Since $V_\mu(x)$ is already fixed at each step, eq.~\eqref{eq:prelink-recursion}
is a left translation of the Haar-distributed $U_\mu(x)$ by a fixed
element, so $dU_\mu(x)=dV_\mu(x+\muhat)$. This recursive relation
indicates that the prelink variables are defined over the minimal
extended lattice $\tilde\Lambda_\text{minimal}$, unlike the links that
are defined over $\Lambda$; see the left panel of
fig.~\ref{fig:v_prelinks}. Iterating along every line replaces each
link integral in eq.~\eqref{eq:Z-links} by a prelink integral over
$\tilde\Lambda_\text{minimal}$, with no additional Jacobian term.
Therefore, the partition function in eq.~\eqref{eq:Z-links} can be
expressed as
\begin{equation}
  \mathcal{Z} = \int \prod_{\mu}  \prod_{x\in\tilde\Lambda_\text{minimal}} dV_\mu(x)\;e^{-S[U]},
  \label{eq:Z-prelinks}
\end{equation}
where the action is a function of prelinks through
eq.~\eqref{eq:U-from-V}. Substituting eq.~\eqref{eq:U-from-V} into the
plaquette formula~\eqref{eq:plaq:U} yields
\begin{equation}
  P_{\mu\nu}(x)
  = V_\mu^\dagger(x)\,V_\mu(x+\muhat)
    \,V_\nu^\dagger(x+\muhat)\,V_\nu(x+\muhat+\nuhat)
    \,V_\mu^\dagger(x+\muhat+\nuhat)\,V_\mu(x+\nuhat)
    \,V_\nu^\dagger(x+\nuhat)\,V_\nu(x),
  \label{eq:plaq-V}
\end{equation}
with eight prelinks entering each plaquette.

With the prelink formulation, the number of apparent degrees of
freedom is increased, but the symmetry is correspondingly enlarged to
account for the added redundancy. The action is invariant under two
distinct transformations, acting on $V_\mu$ from the right and from
the left. The first is local, acting from the right,
\begin{equation}
  V_\mu(x) \to V_\mu(x)\,\Omega^\dagger(x),
  \qquad \Omega(x)\in\SU(N)~\&~\Omega(x+L_\mu\muhat) = \Omega(x).
  \label{eq:prelink-gauge}
\end{equation}
This is exactly equivalent to the ordinary link gauge transformation,
eq.~\eqref{eq:link-gauge}. The second transformation is semi-global,
acting from the left by a factor shared by all prelinks along the
whole line at fixed transverse position,
\begin{equation}
  V_\mu(x)\to Q([x]_\mu)\,V_\mu(x),
  \label{eq:V-left}
\end{equation}
where $[x]_\mu$ denotes the projection of $x$ onto the hyperplane
$x_\mu=0$, obtained by setting $x$'s $\mu$-th coordinate to zero. This
is precisely the freedom, introduced in eq.~\eqref{eq:trivial-insert},
to choose the initial value $V_\mu(x)$ differently for each
$x\in\Lambda_{\perp,\mu}$. We call eq.~\eqref{eq:V-left} the
\emph{initial-value symmetry}, deliberately not a ``gauge'' symmetry,
as it is a redundancy of the prelink parametrization alone, invisible
to the link formulation entirely.

To complete the formulation of the lattice gauge theory with prelink
variables we still need to address the boundary conditions. Recall
that the independent prelinks are defined only on
$\tilde\Lambda_\text{minimal}$: each $V_\mu$ is extended one site
beyond $\Lambda$ in its own direction $\mu$, and not at all in the
others. Defining the boundary condition means specifying $V_\mu(x)$
for $x\in\tilde\Lambda\setminus\tilde\Lambda_\text{minimal}$.
Interestingly, although we assume periodic boundary conditions for the
link variables, the prelink variables can admit a twisted-like
boundary condition,
\begin{equation}
    V_\mu(x + L_\nu \nuhat) = Q([x + L_\nu \nuhat]_\mu)\, V_\mu(x),
    \qquad Q\in\SU(N),
    \qquad \forall\,\mu,\ \nu\ne\mu,
  \label{eq:twisted-bc}
\end{equation}
where $Q$ is the same for every site along the $\muhat$ direction,
since it depends on $x$ only through the projection $[x]_\mu$. Note
that the presence of $Q$ here is exactly the initial-value symmetry of
eq.~\eqref{eq:V-left}, and it is exactly the freedom we exploit in
sec.~\ref{sec:corner-resolution} to resolve the corner constraint.

With this boundary condition, the prelinks are defined on the entire
extended lattice $\tilde\Lambda$. This concludes the formulation of
the lattice gauge theory with prelink variables.

We conclude this subsection with a discussion of fixing the
semi-global initial-value symmetry to a global one. Here we introduce
one specific scheme that gives a simple geometric picture and
motivates the term \emph{holonomy}, introduced in the next subsection;
we first discuss it in the 2D case and then comment on its
generalization to higher dimensions. In two dimensions we set $Q([x]_\mu)$ in eq.~\eqref{eq:V-left} to
\begin{equation}
  Q(m,0) = V_0(m,0),\quad
  Q(0,n) = V_1(0,n).
\label{eq:Q-fix}
\end{equation}
where we remind the reader that
\begin{equation}
  V_0(m,0) = V_0(0,0) \prod_{i=0}^{m-1} U_0(i,0),
  \quad
  V_1(0,n) = V_1(0,0) \prod_{j=0}^{n-1} U_1(0,j).
\end{equation}
With this scheme, both $V_0$ and $V_1$ can be regarded as a product of
links along an L-shaped path starting from the origin, as shown in
the right panel of fig.~\ref{fig:v_prelinks}. Furthermore, we also set
$V_1(0,0)=V_0(0,0)$ at the origin. With this, the semi-global
initial-value symmetry reduces to a global one, associated with
$V_0(0,0)$; one has the freedom to set $V_0(0,0)$ to the identity too,
if needed. A generalization of this scheme to higher dimensions is not
unique but is straightforward, though beyond the scope of this paper.

In addition to fixing the semi-global initial-value symmetry above,
one can impose a scheme to fix the local gauge symmetry too,
consistently with it.
For instance, let us explain the temporal gauge. Using
up the symmetry of eq.~\eqref{eq:prelink-gauge} to set $V_0(x)=I$ for
every $x$ in the physical lattice $\Lambda$, which fixes the ordinary
link temporal gauge $U_0(x)=I$ via eq.~\eqref{eq:U-from-V}; $V_0$
outside $\Lambda$ need not be $I$. In particular, there is a difference between the
minimal extended lattice $\tilde\Lambda_\text{minimal}$ and the full
extended lattice $\tilde\Lambda$: on the points of
$\tilde\Lambda\setminus\tilde\Lambda_\text{minimal}$, eq.~\eqref{eq:twisted-bc} instead fixes $V_0(x)$ to
the constant $Q$ associated with the twisted boundary condition we
choose.

\subsection{Extended lattice and prelink holonomies}
\label{sec:holonomy}

\begin{figure}[tbp]
  \centering
  \includegraphics[width=1\linewidth, trim=0cm 0cm 0cm 1cm, clip]{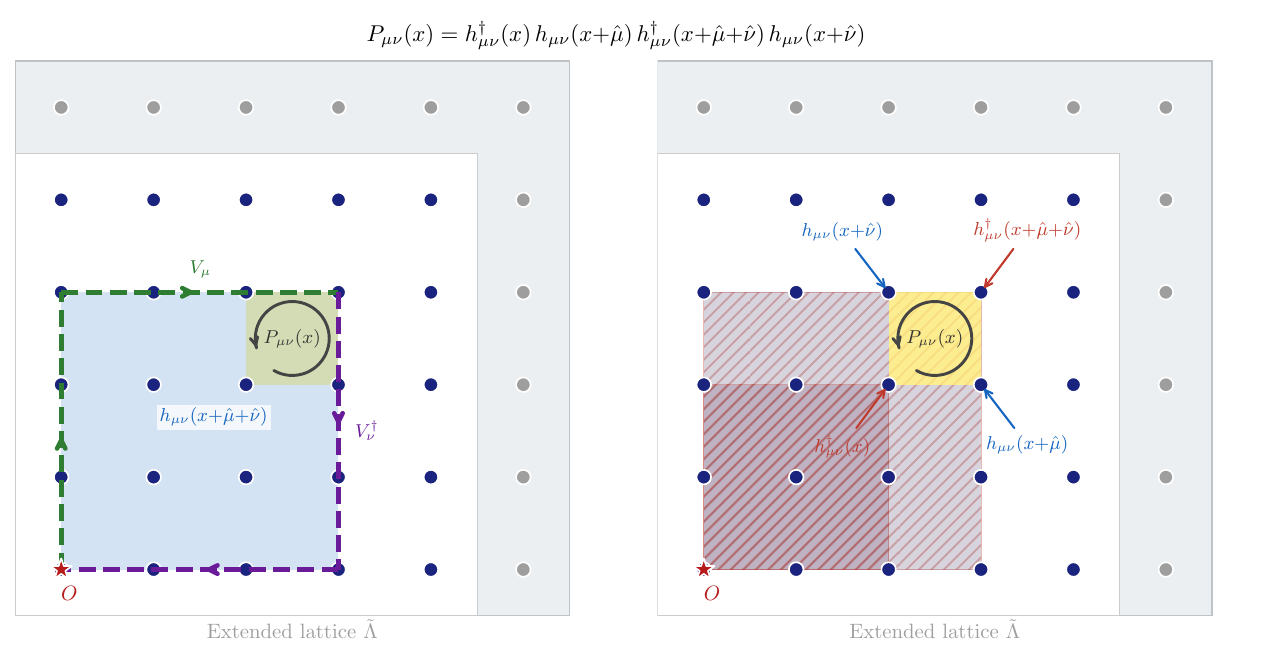}
  \caption{Holonomy $h_{\mu\nu}(x)=V_\mu(x)V_\nu^\dagger(x)$ and the
    plaquette decomposition, eq.~\eqref{eq:plaq-h-general}: the four
    corner rectangles combine so that shared interior Wilson-line
    segments cancel, leaving only the plaquette boundary.}
  \label{fig:h_holonomy}
\end{figure}

The \emph{holonomy} combines two prelinks at the same site,
\begin{equation}
  h_{\mu\nu}(x) = V_\mu(x)V_\nu^\dagger(x),
  \qquad x\in\tilde\Lambda\,.
  \label{eq:holonomy-def}
\end{equation}
defined on the extended lattice.
It also satisfies the Hermitian property $h_{\mu\nu}=h_{\nu\mu}^\dagger$
and the factorization property $h_{\mu\nu}=h_{\mu0}h_{\nu0}^\dagger$,
so the $n-1$ objects $h_{\mu0}$, $\mu=1,\dots,n-1$, carry all the
information.

Because both factors of $h_{\mu\nu}(x)$ are evaluated at the same site
$x$, the right symmetry of eq.~\eqref{eq:prelink-gauge} cancels
identically, indicating the holonomy is \emph{fully} gauge invariant.
Under the left, initial-value symmetry it transforms covariantly,
\begin{equation}
  h_{\mu\nu}(x)\to Q([x]_\mu)\,h_{\mu\nu}(x)\,Q^\dagger([x]_\nu).
  \label{eq:h-left-transform}
\end{equation}
Fixing the semi-global initial-value symmetry to a global one as in
eq.~\eqref{eq:Q-fix}, $h$ becomes a genuine holonomy, as shown in the
left panel of fig.~\ref{fig:h_holonomy}, and transforms as
$h_{\mu\nu}(x)\to Q\,h_{\mu\nu}(x)\,Q^\dagger$ globally. This
construction is essentially the same as the \emph{corner} variables of
refs.~\cite{Bars:1978ry,Batrouni:1984rb}; we use different terminology
because, once the semi-global freedom is fixed, the construction
admits this interesting geometrical description. This fixing is \emph{not} mandatory, however, and we do not
assume it here unless otherwise stated.

One can formulate the Wilson gauge action in terms of
holonomy variables. To this end, one can
define a new operator, different from the original plaquette but
with identical trace,
\begin{align}
  \Tr P_{\mu\nu}(x) &= \Tr \tilde P_{\mu\nu}(x),\\
  \tilde P_{\mu\nu}(x)
  &= h_{\mu\nu}^\dagger(x)\,
    h_{\mu\nu}(x+\muhat)\,
    h_{\mu\nu}^\dagger(x+\muhat+\nuhat)\,
    h_{\mu\nu}(x+\nuhat).
  \label{eq:plaq-h-general}
\end{align}
However, unlike the link and prelink formulations, the holonomy
variables must satisfy some constraints, which complicates the
formulation; interested readers can consult
refs.~\cite{Bars:1978ry,Batrouni:1984rb}. Here we only mention the one
constraint that holds in two dimensions.

\begin{figure}[tbp]
  \centering
  \includegraphics[width=0.55\linewidth, trim=3cm 8.35cm 3cm 3.25cm, clip]{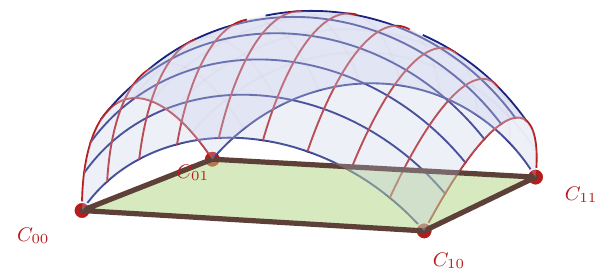}
  \caption{Geometric picture of the corner
    constraint~\eqref{eq:corner-constraint}.}
  \label{fig:h_corner_constraint}
\end{figure}

If we use periodic boundary conditions for the prelinks, as opposed to
the twisted one introduced in eq.~\eqref{eq:twisted-bc}, one can
easily show that, in two dimensions, the holonomy variables are
independent degrees of freedom at all sites of the extended lattice
$\tilde\Lambda$ except at $(L_0,L_1)$. This is easy to see in the
\emph{temporal} gauge with periodic BC, where $V_0(x)=I$ for $x_0<L_0$, but the result
is independent of the gauge fixing. In fact, one can algebraically
show that periodicity is consistent only if the four \emph{corner}
holonomies of $\tilde\Lambda$,
\begin{equation}
  C_{00}=h_{10}(0,0),\quad C_{01}=h_{10}(0,L_1),\quad
  C_{10}=h_{10}(L_0,0),\quad C_{11}=h_{10}(L_0,L_1),
  \label{eq:corners}
\end{equation}
satisfy
\begin{equation} C_{00}\,C_{01}^\dagger\,C_{11}\,C_{10}^\dagger \;=\; I.
  \label{eq:corner-constraint}
\end{equation}
This constraint implies that not all four corner holonomies are
independent.
The holonomy parametrization in 2D is indeed an open-boundary parametrization up to this corner constraint as shown in fig.~\ref{fig:h_corner_constraint}.

Batrouni and Halpern derive the constraint on the finite
periodic lattice, in their string/corner language
\cite[eq.~(4.4)]{Batrouni:1984rb}. They impose it as a hard
delta function on the corner, 
and they do not provide an inverse map back to link/prelink data.
In the next section we address this constraint for the $\SU(N)$ gauge theory in two dimensions.

\section{Open-boundary holonomy parametrization}
\label{sec:constraint}

In the above section we discussed the holonomy parametrization of gauge theories and argued that, even in the 2D case, it requires a constraint connecting the four corners of the extended lattice $\tilde\Lambda$ under periodic boundary conditions.
In this section we trade this hard constraint for a statistical sampling problem by factorizing it into a boundary-free theory and a conditional sampling term that will be handled using importance sampling. We then show that the plaquettes constructed from the holonomies can be sampled independently and that changes in the plaquettes can be pushed to the holonomies. We also discuss how the holonomies can be mapped to gauge prelinks and, in the end, to gauge links, with the resulting configurations reweighted to match the exact measure.

\subsection{The corner constraint as a group-commutator problem}
\label{sec:corner-resolution}

The corner constraint~\eqref{eq:corner-constraint}, which originates
from imposing periodic boundary conditions on prelinks, makes it hard
to sample holonomies. The constraint can be relaxed if we
use the twisted BC of eq.~\eqref{eq:twisted-bc}.

\begin{figure}[t]
  \centering
  \includegraphics[width=\linewidth]{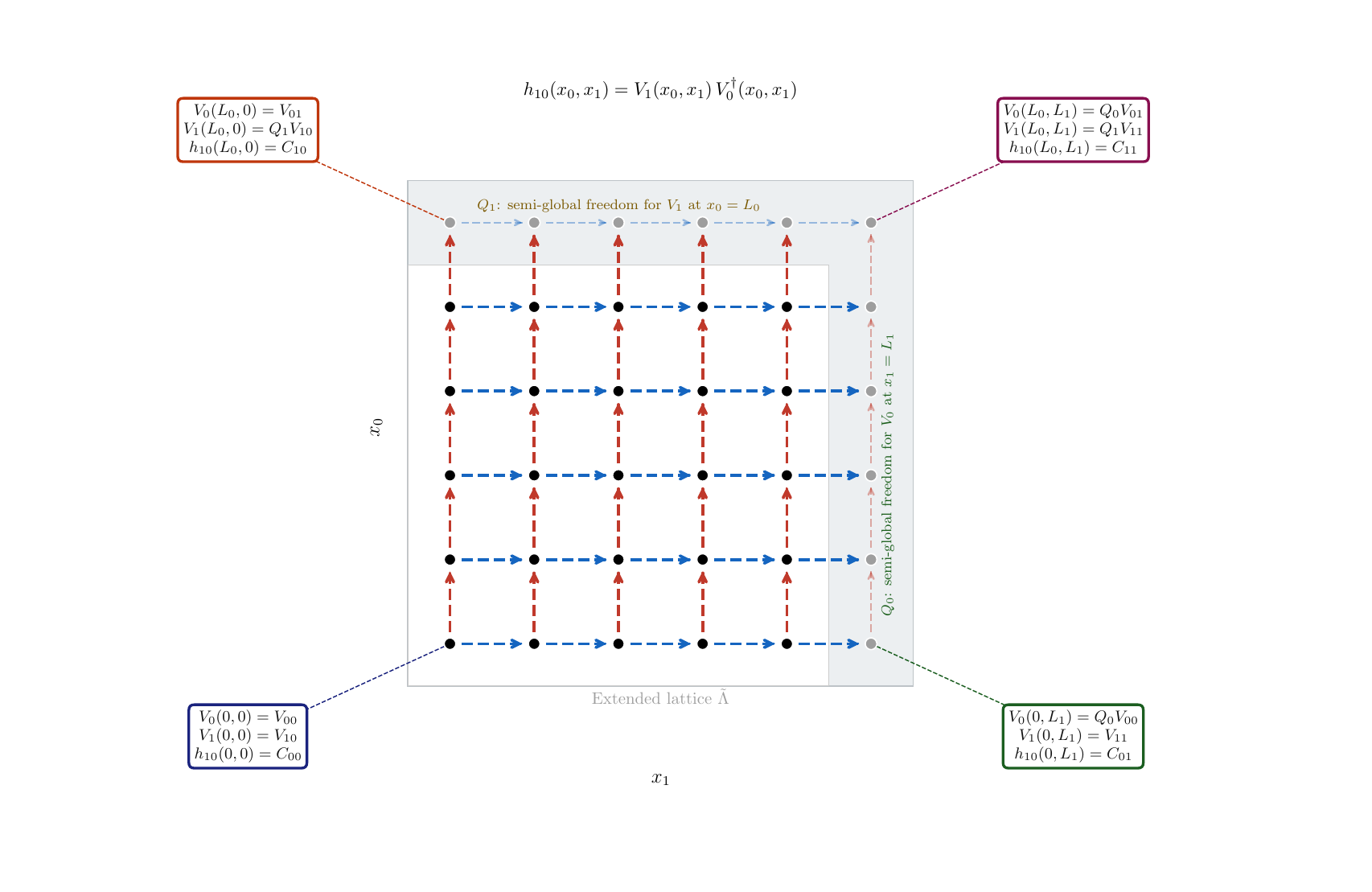}
  \caption{Corner labeling used to derive
    eqs.~\eqref{eq:corner-C-def}--\eqref{eq:commutator}: seed prelinks $V_{01},V_{11}$ and
    semi-global freedoms $Q_0$ (right column, $x_1=L_1$) and $Q_1$ (top row,
    $x_0=L_0$).}
  \label{fig:corner-prelinks}
\end{figure}

Before we give a precise holonomy formulation, it would be worthwhile to see how it is affected by the prelinks boundary condition.
Figure~\ref{fig:corner-prelinks} shows the prelinks on the extended 2D lattice
with the twisted boundary condition.
The prelinks at the origin, $V_{00}\equiv V_0(0,0)$ and $V_{10}\equiv V_1(0,0)$,
together with the twisted boundary condition terms $Q_0$ (acting on $V_0$ at
$x_1=L_1$) and $Q_1$ (acting on $V_1$ at $x_0=L_0$), determine the four
corner holonomies:
\begin{equation}
  C_{00} = V_{10}\,V_{00}^\dagger,\quad
  C_{01} = V_{11}\,(Q_0V_{00})^\dagger,\quad
  C_{10} = Q_1V_{10}\,V_{01}^\dagger,\quad
  C_{11} = Q_1V_{11}\,V_{01}^\dagger Q_0^\dagger,
  \label{eq:corner-C-def}
\end{equation}
where $V_{01}\equiv V_0(L_0,0)$ and $V_{11}\equiv V_1(0,L_1)$ are the remaining
independent prelinks.
Now define the corner products
\begin{align}
  C &\equiv C_{00}\,C_{01}^\dagger\,C_{11}\,C_{10}^\dagger\,,
  \\
  Z &\equiv C_{00}\,C_{10}^\dagger\,C_{11}\,C_{01}^\dagger\,.
  \label{eq:corner-products}
\end{align}
Under periodic BC on prelinks, i.e., $Q_0=Q_1=I$, we have $C=I$, which is the same corner constraint
eq.~\eqref{eq:corner-constraint}. To avoid imposing this constraint
one may instead treat $Q_0$ and $Q_1$ as free parameters and integrate over them, which is the approach taken in the rest of this section for
$\SU(N)$ gauge theory.
Moreover, note that $Z$ depends only on the corner data
$\{C_{00},C_{01},C_{10},C_{11}\}$;
neither $Q_0$ nor $Q_1$ appears in
its definition, so it is independent of the prelinks BC.

Given holonomy matrices $\{C_{00}, C_{01}, C_{10}, C_{11}\}$, our aim
is to compute the prelink matrices $\{V_{00}, V_{10}, V_{01}, V_{11}\}$ such that the
resulting configuration reproduces the same action.
We express eq.~\eqref{eq:corner-C-def} as
\begin{equation}
  V_{10} = C_{00} V_{00},\quad
  V_{11} = C_{01} Q_0 V_{00},\quad
  V_{01} = C_{10}^\dagger Q_1 C_{00} V_{00},\quad
  C_{11} = Q_1 C_{01} Q_0 C_{00}^\dagger Q_1^\dagger C_{10} Q_0^\dagger.
  \label{eq:prelink-corner-system}
\end{equation}
Using the temporal gauge, in which $V_{00} = I$, the first three equations give $\{V_{10}, V_{01}, V_{11}\}$ as functions of
$\{C_{00}, C_{01}, C_{10}, Q_0, Q_1\}$,
where $\{Q_0, Q_1\}$ themselves must be chosen such that the fourth equation holds.
The change of variables
$Q_1 = C_{10} C_{00}^\dagger X$, $Q_0 = C_{01}^\dagger Y C_{00}$,
turns the fourth equation into the \emph{group commutator equation}
\begin{equation}
  Z = X\,Y\,X^\dagger\,Y^\dagger\,,
  \label{eq:commutator}
\end{equation}
with $Z$ as defined in eq.~\eqref{eq:corner-products}, for
$X,Y\in\SU(N)$. Once $X,Y$ are known, the twisted BC terms $Q_0$
and $Q_1$ are
fixed via the formulas above, and, in turn, $\{V_{00},V_{10},V_{01},V_{11}\}$
are determined.

We now derive the holonomy formulation directly from the prelink
partition function, rather than simply asserting it as in
sec.~\ref{sec:holonomy}.
Starting from the prelink partition function~\eqref{eq:Z-prelinks}, and trivially integrating over the
twisted-BC terms $Q_0, Q_1$ exactly as in eq.~\eqref{eq:trivial-insert},
the measure reads
\begin{equation}
dQ_0\,dQ_1\!\!\prod_{\mu=0,1}\,\prod_{x\in\tilde\Lambda_\text{minimal}}\!\!
  dV_\mu(x)\; e^{-S[V]}\,.
  \label{eq:prelink-QQ-measure}
\end{equation}
Representing this measure in terms of holonomy variables does not
require fixing the gauge. However, we do so because
it also shows how to map the holonomies back to the prelinks. In the
temporal gauge, $V_0(x)=I$ on the physical lattice $\Lambda$
(sec.~\ref{sec:params}), so we can simply define
\begin{equation}
  h(x_0,x_1) =
  \begin{cases}
    V_1(x_0,x_1), & x_0<L_0,\; x_1<L_1,\\[4pt]
    V_1(x_0,L_1)\,Q_0^\dagger, & x_0<L_0,\;x_1=L_1,\\[4pt]
    Q_1\,V_1(0, x_1)\,V_0^\dagger(L_0,x_1), & x_0=L_0,\; x_1<L_1.
  \end{cases}
  \label{eq:prelink-h-def}
\end{equation}
With this change of variables, $h(x_0,x_1)$ is defined at every point
of the extended lattice except $(L_0,L_1)$.
With this, eq.~\eqref{eq:prelink-QQ-measure} can therefore be written as
\begin{equation}
 dQ_0\, dQ_1 \prod_{x\in\tilde\Lambda\setminus\{(L_0, L_1)\}}
  dh(x)\; e^{-S[h]}\,,
  \label{eq:holonomy-QQ-measure}
\end{equation}
with the action depending on $h$ only through the plaquette
combination already given in eq.~\eqref{eq:plaq-h-general}.
We then change $(Q_0, Q_1)$ to $(Y, X)$ using the transformation given above, which
preserves the volume, and then use the group-commutator encoder of
sec.~\ref{sec:commutator-encoder-decoder} to write
\begin{equation}
 dX\, dY = dD\, d\Omega\, dZ\, p_\text{encode}(D, \Omega, Z)\,.
\end{equation}
We then change the variable from $Z$ to $C_{11}$, which is volume preserving at fixed
$C_{00},C_{01},C_{10}$. 
Since $C_{11}=h(L_0,L_1)$, it folds back into the product above
and restores the full extended lattice, the equation reads
\begin{equation}
 dD\,d\Omega\, p_\text{encode}(D, \Omega, Z(h)) \prod_{x\in\tilde\Lambda}
  dh(x)\; e^{-S[h]}\,.
  \label{eq:holonomy-encoded-measure}
\end{equation}

Equation~\eqref{eq:holonomy-encoded-measure} factorizes the problem into two parts.
The first part is the open-boundary holonomy theory, which itself can be fully
factorized and sampled independently, plaquette by plaquette. The second part is a conditional term, depending on $Z(h)$, for sampling $(D, \Omega)$.

\subsection{Group-commutator encoding and sampling \texorpdfstring{$(X,Y)$}{(X,Y)}}
\label{sec:commutator-sampling}

Given $Z$, we need $(X,Y)$ drawn from the conditional law that the
$\SU(N)\times\SU(N)$ Haar measure induces under eq.~\eqref{eq:commutator}.
We parametrize $(X,Y)$ by an \emph{encoder} that makes $Z$ one of the
coordinates, sample the remaining coordinates given $Z$, and use a
\emph{decoder} to construct $X$ and $Y$ from them. We do this for $\SU(2)$ and
$\SU(3)$.

\subsubsection{Encoder and decoder}
\label{sec:commutator-encoder-decoder}

Assuming that the eigenvalues of $Y$ are non-degenerate and $Z\neq I$, we
introduce an encoder $(X,Y)\to(D,\Omega,Z)$ and a
decoder $(D,\Omega,Z)\to(X,Y)$ for $\SU(2)$ and $\SU(3)$.
Let $\Omega$ be the matrix of eigenvectors of $Y$, so that
$Y=\Omega\Lambda\Omega^\dagger$ with $\Lambda=\operatorname{diag}(\lambda_k)$.
Since $ZY=XYX^\dagger$, the matrix $X\Omega$ also diagonalizes $ZY$, with the
same eigenvalues $\Lambda$. So if $P$ is the matrix of eigenvectors of $ZY$,
ordered like $\Lambda$, then $D=P^\dagger X\Omega$ is a diagonal $\SU(N)$ matrix.
The encoder and decoder are
\begin{align}
  \textbf{encode:}\quad&
  Z = XYX^\dagger Y^\dagger,\quad
  Y = \Omega\Lambda \Omega^\dagger,\quad
  D = P^\dagger X \Omega,
  \label{eq:encoder}
  \\
  \textbf{decode:}\quad&
  \Lambda:\ \Tr\bigl[(I-\Omega^\dagger Z\Omega)\Lambda\bigr]=0,\quad
  Y=\Omega\Lambda \Omega^\dagger,\quad
  X = P D \Omega^\dagger.
  \label{eq:decoder}
\end{align}
Here $\Omega$ and $P$ are projected to $\SU(N)$ and use a fixed ordering of the
eigenvalues.

The equation for $\Lambda$ in the decoder is the equality of the traces of
$ZY$ and $Y$. For $\SU(2)$ and $\SU(3)$ it is equivalent to
$\operatorname{spectrum}(ZY)=\operatorname{spectrum}(Y)$, which determines
$\Lambda$ up to discrete choices. 
For $\SU(2)$ it reduces to $\mathrm{Re}(m_1\lambda_1)=0$ with
$m=\operatorname{diag}(I-\Omega^\dagger Z\Omega)$, fixing $\lambda_1$ up to the
$\SU(2)$ center ($\pm\Lambda$). For $\SU(3)$ it is a triangle closure on the
three $|m_k|$, fixing $\Lambda$ up to $6$ choices (3 center rotations times 2
chiralities). We select one by sorting the eigen-angles consistently with the
encoder. For $N\ge4$ equal traces are only necessary, which is why the decoder is
restricted to $\SU(2)$ and $\SU(3)$.
The decoder also requires $Z\neq I$ because for $Z=I$ we have
$ZY=Y$, the equation holds for any $\Lambda$, and $\Lambda$ is not determined.

Decoding an encoded pair returns the original pair $(X,Y)$. The converse need not
hold; encoding a decoded triple may change the phases of $\Omega$ and $D$, as
when eigenvectors are recomputed from a composed matrix. We assumed above that $Y$
has non-degenerate eigenvalues. Otherwise, $D$ is not diagonal and the decoder
above fails. Matrices $Y$ with degenerate eigenvalues form a set of measure zero.

\subsubsection{Jacobian: the spectral twist}
\label{sec:spectral-twist}

The density $p_\text{encode}$ of eq.~\eqref{eq:holonomy-encoded-measure} is the
inverse of the Jacobian of the encoder. This Jacobian reduces to a more
elementary problem. For $G\in\SU(N)$, write
$G=W\Lambda_GW^\dagger$ with the eigenvalues sorted by argument, and define the
\emph{spectral twist}
$\tilde Z\equiv G\Lambda_G^\dagger=W \Lambda_G W^\dagger\Lambda_G^\dagger$, a group commutator with one
factor diagonal.
The Jacobian depends only on the matrix of eigenvectors $W$, through the
doubly-stochastic matrix $B_{ij}\equiv|W_{ij}|^2$,
\begin{equation}
  \left|\det\frac{d\tilde Z}{dG}\right|
  = \det(I-B)\Big|_{\{\eta:\sum_m \eta_m=0\}}.
  \label{eq:spectral-twist-jacobian}
\end{equation}
To see this, let $A=G^\dagger dG\in\su(N)$. Then
$\tilde Z^\dagger d\tilde Z=\Lambda_GA\Lambda_G^\dagger-\Lambda_G^\dagger d\Lambda_G$,
where $\Lambda_G^\dagger d\Lambda_G$ is the diagonal matrix with entries
$(W^\dagger AW)_{kk}$. On the
off-diagonal entries of $A$ this map multiplies by the phases
$\lambda_i\bar\lambda_j$. On the traceless diagonal entries,
$A=i\operatorname{diag}(\eta)$ with $\sum_m \eta_m=0$, it acts as
$\eta\mapsto(I-B^{\mathsf T})\eta$, because $(W^\dagger AW)_{kk}=i\sum_j B_{jk}\eta_j$.
The map is block triangular, which gives eq.~\eqref{eq:spectral-twist-jacobian}.
For $N=2$ and $N=3$, we have
\begin{align}
  \SU(2):\quad & \left|\det\frac{d\tilde Z}{dG}\right| = 2-\Tr B = 2|W_{12}|^2,
  \label{eq:spectral-twist-su2}
  \\
  \SU(3):\quad & \left|\det\frac{d\tilde Z}{dG}\right|
  = 2 - \Tr B + \det B.
  \label{eq:spectral-twist-su3}
\end{align}
The bridge to sec.~\ref{sec:commutator-encoder-decoder} is conjugation by the
matrix of eigenvectors of $Y$, turning $Z$ into
$\tilde Z=\Omega^\dagger Z\Omega$, which is the spectral twist of
$G=\Omega^\dagger X \Omega \Lambda \Omega^\dagger X^\dagger\Omega$. The matrix of eigenvectors of this $G$ is $W=\Omega^\dagger X\Omega$. The Jacobian of the
encoder is exactly $|\det(d\tilde Z/dG)|$, so
\begin{equation}
  p_\text{encode}(D,\Omega,Z) = \frac{1}{|\det(d\tilde Z/dG)|},
  \label{eq:p-encode}
\end{equation}
with eqs.~\eqref{eq:spectral-twist-su2} and~\eqref{eq:spectral-twist-su3} giving
it in closed form for both groups.

\subsubsection{The conditional law and sampling with Euler angles}
\label{sec:commutator-euler}
\label{sec:commutator-conditional}

The encoded density factorizes as
$p_\text{encode}(D,\Omega,Z)=\rho(Z)\,P(D,\Omega\mid Z)$, where $\rho(Z)$ is the
density of $Z$ (sec.~\ref{sec:commutator-density}) and $P(D,\Omega\mid Z)$ is the
conditional law that we need to sample.
There are two structural facts about $p_\text{encode}$ that allow us to simplify
the problem.
First, the density does not depend on $D$, so $D$ is uniform and independent of $Z$.
Second, writing $Z$ as
$Z=\Omega_Z\operatorname{diag}(e^{i\phi_1},\dots,e^{i\phi_N})\,\Omega_Z^\dagger$
and defining the \emph{relative frame}
\begin{equation}
  R \equiv \Omega_Z^\dagger\,\Omega \in \SU(N),
  \label{eq:relative-frame}
\end{equation}
we can parametrize the dependence of $\Omega$ on $Z$.
Under a simultaneous conjugation $(X,Y)\to(gXg^\dagger,gYg^\dagger)$, we have $Z\to gZg^\dagger$, $\Omega\to g\Omega$, and
$\Omega_Z\to g\Omega_Z$, while $R$ and $D$ are invariant.
The conditional sampling of $\Omega$ given $Z$ reduces to sampling $R$ given
the eigen-angles of $Z$.

We sample $R$ in Euler-angle coordinates. For $\SU(2)$,
$R=D(\varphi)\,R(\vartheta)\,D(\psi)$, and for $\SU(3)$,
\begin{equation}
  R = D_1(\alpha_1,\alpha_2)\,R_{23}(\theta_{23})\,
      U_{13}(\theta_{13},\delta)\,R_{12}(\theta_{12})\,D_2(\beta_1,\beta_2),
  \label{eq:su3-euler}
\end{equation}
with $D_{1,2}$ diagonal $\SU(3)$ matrices, $R_{12},R_{23}$ real rotations in the
$(1,2)$ and $(2,3)$ planes, and $U_{13}$ a rotation in the $(1,3)$ plane carrying
the only complex phase (the four central parameters are those of the CKM
parametrization). We use the \emph{uniform} version of these coordinates,
obtained by a monotonic reparametrization of each angle such that all
coordinates are exactly $\mathrm{Uniform}(0,1)$ under the Haar measure. The Haar
measure is then flat on $[0,1]^3$ for $\SU(2)$ and on $[0,1]^8$ for $\SU(3)$.
Conditioning on $Z$ deforms only the central coordinates ($\vartheta$ for
$\SU(2)$, and $(\theta_{23},\theta_{13},\delta,\theta_{12})$ for $\SU(3)$), while
the outer phases remain uniform.

The sampling procedure is the same for both groups.
We draw $D$ Haar over the diagonal $\SU(N)$ matrices and $R_0$ Haar on
$\SU(N)$, transform the central coordinates of $R_0$ by a map $T_Z$ that depends
on $Z$ through its eigen-angles, giving $R=T_Z(R_0)$, set $\Omega=\Omega_Z R$,
and decode, $(X,Y)=\text{decode}(D,\Omega,Z)$.
Let $q(\Omega\mid Z)$ be the density of the resulting $\Omega$.
The importance weight of the conditional sampler is $w=p/q$, where
$p(\Omega\mid Z)=1/\bigl(\rho(Z)\,|\det(d\tilde Z/dG)|\bigr)$ is known in closed
form, from eqs.~\eqref{eq:spectral-twist-su2} and~\eqref{eq:spectral-twist-su3}
and eq.~\eqref{eq:commutator-density}. If $j_T$ is the Jacobian of $T_Z$ in the
uniform coordinates, then $q=1/j_T$ and
\begin{equation}
  \log w = -\log\bigl|\det(d\tilde Z/dG)\bigr| - \log\rho(Z) + \log j_T.
  \label{eq:conditional-importance-weight}
\end{equation}
The sampling problem is thus reduced to finding the map $T_Z$, which is where
the two groups differ.

\subsubsection{Closed-form sampler for \texorpdfstring{$\SU(2)$}{SU(2)}}
\label{sec:commutator-sampling-su2}

For $\SU(2)$, the relative frame $R$ has three uniform coordinates $(a,b,c)$,
where $b=|R_{00}|^2$ is the one in the middle. Conditioning on $Z$ deforms only
$b$, while $a$ and $c$ stay uniform. The conditional law of $b$ has a closed-form
cumulative distribution function,
\begin{equation}
  F(b\mid\theta) = \frac{\arctan\bigl[(2b-1)\cot(\theta/2)\bigr] + (\pi-\theta)/2}
  {\pi-\theta},
  \label{eq:su2-conditional-cdf}
\end{equation}
where $e^{\pm i\theta}$, with $0<\theta<\pi$, are the eigenvalues of $Z$.
We draw $s\sim\mathrm{Uniform}(0,1)$ and set $b=F^{-1}(s\mid\theta)$, that is,
$b=\tfrac12\bigl[1+\tan(\theta/2)\tan\bigl((\pi-\theta)(s-\tfrac12)\bigr)\bigr]$.
We then construct $R$, and in turn $\Omega$. This reproduces the target
conditional law exactly, so the sampler is exact and $w=1$ in
eq.~\eqref{eq:conditional-importance-weight}.

\subsubsection{Trained sampler for \texorpdfstring{$\SU(3)$}{SU(3)}}
\label{sec:commutator-sampling-su3}

For $\SU(3)$, a closed-form conditional sampler is not yet available, and we are
still working on deriving one. For now, the map $T_Z$ is a trained normalizing
flow. It acts on the four central coordinates of $R$, and it is conditioned on the
real and imaginary parts of the eigenvalues of $Z$. The four outer phases and $D$
remain uniform. The flow is a masked autoregressive rational-quadratic spline flow
with four spline segments and two hidden layers of width $32$.

We train the flow by self-learning. The prior consists of $D$, $\Omega$, and $Z$,
all drawn from the Haar measure. A flow transforms $Z$, a second flow
transforms $\Omega$ given $Z$, and the decoder turns $(D,\Omega,Z)$ into $(X,Y)$.
The target is the Haar distribution of $(X,Y)$. We minimize the reverse
KL divergence, with path-gradient estimators~\cite{vaitl:2022path}, between the distribution of the
generated pair and this target. After training, only the conditional part
$q(\Omega\mid Z)$ is used, with $Z$ supplied by the holonomies. Training uses
$50\,000$ epochs with batch size $1024$, learning rate $0.002$, weight decay
$0.01$, and a cosine-annealed learning rate.

Because $q$ is only a trained approximation, the weight $w$ of
eq.~\eqref{eq:conditional-importance-weight} is not one. It can be computed
exactly, since $\rho(Z)$ and the Jacobian of the spectral twist are known in closed
form for $N=3$, and $\log j_T$ is the log-Jacobian of the flow. 
At fixed $Z$, the effective sample size of this correction is empirically $0.98$.

\subsubsection{Marginal distribution of the group commutator \texorpdfstring{$Z$}{Z}}
\label{sec:commutator-density}

For $X,Y\in\SU(N)$ independently Haar-distributed, $Z=XYX^\dagger Y^\dagger$
has a density $\rho(Z)$ with respect to the Haar measure $dZ$,
\begin{equation}
  \rho(Z) = \int dX\,dY\;\delta_{\SU(N)}(Z^\dagger XYX^\dagger Y^\dagger)
  = \int p_\text{encode}(D,\Omega,Z)\,dD\,d\Omega \,.
  \label{eq:commutator-density-def}
\end{equation}
The first form is a classical
integral; expanding the group delta function in characters and integrating
over $X$ and $Y$ gives
\begin{equation}
  \rho(Z)=\sum_R \frac{\chi_R(Z)}{d_R},
  \label{eq:commutator-density-series}
\end{equation}
summed over the irreducible representations $R$ of dimension $d_R$~\cite[Proposition~3.1]{Mulase:2002vol}. We now sum
this series for $\SU(2)$ and $\SU(3)$.

For $\SU(2)$, with $\theta\in[0,\pi]$ the class angle ($Z$ has eigenvalues
$e^{\pm i\theta}$), the $m$-dimensional irreducible representation has
$\chi_m=\sin(m\theta)/\sin\theta$ and $d_m=m$ for $m=1,2,\dots$~\cite[eq.~(10.107) with $m=2j+1$]{Greiner:1994sym}, and the series sums to
\begin{equation}
  \rho(Z) = \frac{1}{\sin\theta}\sum_{m\ge1}\frac{\sin m\theta}{m}
  = \frac{\pi-\theta}{2\sin\theta}.
  \label{eq:commutator-density-su2}
\end{equation}

For $\SU(3)$, let $z_k=e^{i\phi_k}$, $k=1,2,3$, be the eigenvalues of $Z$,
with each $\phi_k\in[0,2\pi)$, so that $z_1z_2z_3=1$ and
$\phi_1+\phi_2+\phi_3\in\{2\pi,4\pi\}$, and let
$\Delta(z)=(z_1-z_2)(z_1-z_3)(z_2-z_3)$.
We also exclude the identity matrix, for which the density diverges as we discuss below.
The irreducible representations $R$ are
labeled by integers $p,q\ge1$.
Their characters and dimensions are~\cite[eq.~(3.1)--(3.4)]{Baaquie:1988su3}
\begin{align}
  \chi_{p,q} &=-\frac{1}{\Delta(z)}\Bigl[e^{i(p\phi_1-q\phi_2)}-e^{i(-q\phi_1+p\phi_2)}
  +e^{-i((p+q)\phi_1+p\phi_2)}\nonumber\\
  &\qquad-e^{-i(p\phi_1+(p+q)\phi_2)}+e^{i(q\phi_1+(p+q)\phi_2)}
  -e^{i((p+q)\phi_1+q\phi_2)}\Bigr],
  \label{eq:su3-character} \\
  d_{p,q} &= p q (p+q) / 2.
\end{align}
Each of the six exponentials is
$e^{i(m\phi_1+n\phi_2)}$, with
\begin{equation}
  (m,n)\in\bigl\{(p,-q),\ (-q,p),\ (-p-q,-p),\ (-p,-p-q),\ (q,p+q),\ (p+q,q)\bigr\},
\end{equation}
and in each case $mn(m-n)/2$ is indeed $d_{p,q}$ times the sign of the corresponding term. As $(p,q)$
runs over all representations, these pairs run over all integers $m,n$ with
$m,n\neq0$ and $m\neq n$, each pair exactly once. Hence
\begin{equation}
  \frac{\rho(Z)\,\Delta(z)}{2}
  = \sum_{\substack{m,n\neq0\\ m\neq n}} \frac{e^{i(m\phi_1+n\phi_2)}}{mn(m-n)}.
  \label{eq:su3-lattice-sum}
\end{equation}

To sum this series, write $m\phi_1+n\phi_2=\tfrac12[(m+n)s+(m-n)u]$, with
$s=\phi_1+\phi_2$ and $u=\phi_1-\phi_2$, and differentiate with respect to $u$ at
fixed $s$. Each term is multiplied by $i(m-n)/2$, which cancels the factor $m-n$
in the denominator, and the restriction $m\neq n$ becomes the subtraction of the
terms with $m=n$:
\begin{equation}
  \partial_u\frac{\rho(Z)\,\Delta(z)}{2}
  = \frac i2\Bigl[\sum_{m\neq0}\frac{e^{im\phi_1}}{m}\sum_{n\neq0}\frac{e^{in\phi_2}}{n}
  -\sum_{m\neq0}\frac{e^{-im\phi_3}}{m^2}\Bigr],
\end{equation}
where $e^{im(\phi_1+\phi_2)}=e^{-im\phi_3}$ has been used. With the standard
Fourier series $\sum_{m\neq0}e^{im\phi}/m=i(\pi-\phi)$ for $0<\phi<2\pi$ and
$\sum_{m\neq0}e^{im\phi}/m^2=\tfrac12(\pi-\phi)^2-\tfrac{\pi^2}{6}$ for
$0\le\phi\le2\pi$,
\begin{equation}
  \partial_u\frac{\rho(Z)\,\Delta(z)}{2}
  = -\frac i2\Bigl[(\pi-\phi_1)(\pi-\phi_2)+\tfrac12(\pi-\phi_3)^2-\tfrac{\pi^2}{6}\Bigr].
\end{equation}
Here $\phi_3=2\pi k-s$, with $k=1$ for $s<2\pi$ and $k=2$ otherwise, and in the
variables $s$ and $u$ the right-hand side equals $\partial_u$ of
$-\tfrac i6V$, where
$V=(\phi_1-\phi_2)(\phi_1-\phi_3)(\phi_2-\phi_3)=\tfrac u4\bigl[(3s-4\pi k)^2-u^2\bigr]$.
Both the series and $V$ change sign under $\phi_1\leftrightarrow\phi_2$, i.e.,
$u\to-u$, and hence vanish at $u=0$, so
\begin{equation}
  \frac{\rho(Z)\,\Delta(z)}{2}=-\frac{i}{6}(\phi_1-\phi_2)(\phi_1-\phi_3)(\phi_2-\phi_3),
\end{equation}
Since
$\Delta(z)=-8i\sin\frac{\phi_1-\phi_2}{2}\sin\frac{\phi_1-\phi_3}{2}\sin\frac{\phi_2-\phi_3}{2}$,
we obtain
\begin{equation}
  \rho(Z) = \frac{1}{N}\prod_{i<j}
  \frac{(\phi_i-\phi_j)/2}{\sin\bigl[(\phi_i-\phi_j)/2\bigr]}
  = \frac{1}{N}\,\frac{|\Delta(\phi)|}{|\Delta(z)|},
  \qquad N=3\,.
  \label{eq:commutator-density}
\end{equation}
Up to the factor $1/N$, the density is the ratio of the Vandermonde determinants
of the eigen-angles and of the eigenvalues. Each factor of the product is
positive, since $|\phi_i-\phi_j|<2\pi$.

Interestingly, eq.~\eqref{eq:commutator-density} holds for $\SU(2)$
too; the eigen-angles in $[0,2\pi)$ are $\phi_1=\theta$ and $\phi_2=2\pi-\theta$.
For both $N=2$ and
$N=3$, $\rho$ diverges as $Z \to I$,
where some eigen-angles approach 0 and others $2\pi$,
and $\rho$ is finite and positive elsewhere.
(At $Z=I$ itself all $\phi_k=0$ and the formula does not apply
because we assumed $\phi_1+\phi_2+\phi_3\in\{2\pi,4\pi\}$.)

For $\SU(2)$, the result can also be obtained directly from the second form of
eq.~\eqref{eq:commutator-density-def}, by integrating $p_\text{encode}$ over $D$
and $\Omega$. Let $R=\Omega_Z^\dagger\Omega$ be the relative frame and $b=|R_{00}|^2$
its modulus coordinate (sec.~\ref{sec:commutator-sampling-su2}), which is
uniformly distributed on $[0,1]$ under the Haar measure. The first diagonal entry
of $\tilde Z=\Omega^\dagger Z\Omega=R^\dagger\operatorname{diag}(e^{i\theta},e^{-i\theta})R$
is then $\cos\theta+i(2b-1)\sin\theta$. With $d$ equal to one minus this entry,
the closure condition $\mathrm{Re}(d\lambda_1)=0$ and
$(\tilde Z\Lambda)_{11}=\sum_j|W_{1j}|^2\lambda_j$ give
$2|W_{12}|^2=|d|^2/\mathrm{Re}\,d$, so eq.~\eqref{eq:spectral-twist-su2} becomes
$|\det(d\tilde Z/dG)|=2\bigl[\sin^2(\theta/2)+(2b-1)^2\cos^2(\theta/2)\bigr]$.
Since $p_\text{encode}$ depends only on $b$,
\begin{equation}
  \rho(Z)=\int_0^1\frac{db}{2\bigl[\sin^2(\theta/2)+(2b-1)^2\cos^2(\theta/2)\bigr]}
  =\frac{\arctan\bigl(\cot(\theta/2)\bigr)}{\sin\theta}
  =\frac{\pi-\theta}{2\sin\theta},
\end{equation}
in agreement with eq.~\eqref{eq:commutator-density-su2}.

\subsection{Sampling workflow}
\label{sec:boundary-free}
\label{sec:workflow}

In the measure of
eq.~\eqref{eq:holonomy-encoded-measure},
$S[h]$ is the open-boundary holonomy action,
\begin{equation}
  S[h] = - \frac{\beta}{N} \sum_{x\in\Lambda}
    \ReTr \left( h_{10}^\dagger(x)\,
    h_{10}(x+\muhat)\,
    h_{10}^\dagger(x+\muhat+\nuhat)\,
    h_{10}(x+\nuhat)\right)\,.
  \label{eq:Z-holonomy}
\end{equation}
The action is a sum over the physical lattice $\Lambda$ while the
holonomies live on the extended lattice $\tilde\Lambda$; in this sense the
simulation is boundary free. The open-boundary holonomy theory
$\prod dh\,e^{-S[h]}$ factorizes into independent plaquettes; it differs from
the exact marginal of $h$ only by the weight that depends on $Z(h)$, which
couples the corners. We therefore sample as follows.

\begin{enumerate}
\item \textbf{Plaquettes.} We train a model for the single-plaquette
  distribution using a spectral-flow model: a uniform (Haar)
  prior on $\SU(N)$ is mapped by a rational-quadratic-spline flow acting on the
  conjugation-invariant (spectral) parameters of the matrix; one parameter for
  $\SU(2)$ and two parameters for $\SU(3)$~\cite{Boyda:2020hsi, Komijani:2025yjz}. The model is trained by minimizing the
  reverse KL divergence to the single-plaquette Wilson weight
  $e^{(\beta/N)\ReTr P}$. All $L_0\times L_1$ plaquettes are then drawn
  independently from the trained model, which also returns the log-density
  $\log q$ of each draw.
\item \textbf{Accept/reject.} Although the effective sample size of these
  models is high (the acceptance rate exceeds $99\%$ in the cases of
  sec.~\ref{sec:simulations}), the plaquette samples still need to be
  corrected, which we do with a Metropolis--Hastings accept/reject step for
  independent proposals, based on the ratio of the target weight to $q$.
  It runs along the sequence of samples, with an independent chain for each
  plaquette position, so that the accepted plaquettes are exact samples of the
  open-boundary theory in the stationary limit.
\item \textbf{Holonomies.} The holonomies are constructed from the
  plaquettes by cumulative matrix products along both directions, with an
  identity prepended. This fixes the holonomies on the first column and first
  row to the identity, using the symmetry $h(x)\to a(x_0)\,h(x)\,b^\dagger(x_1)$
  of the action, which leaves $\rho(Z(h))$ invariant as well. Each step of the
  construction multiplies a plaquette by fixed elements, so the map is
  volume preserving. 
\item \textbf{Corner.} Given $Z(h)$, we sample $(D,\Omega)$ from
  $P(D,\Omega\mid Z(h))$ as described in sec.~\ref{sec:commutator-sampling}
  exactly for $\SU(2)$, with a trained model for $\SU(3)$. We decode them to $(X,Y)$, and obtain $Q_1=C_{10}C_{00}^\dagger X$ and
  $Q_0=C_{01}^\dagger Y C_{00}$. The configuration
  carries the weight $\rho(Z(h))$, multiplied for $\SU(3)$ by the weight $w$ of the conditional sampler.
  
\item \textbf{Prelinks and links.} The holonomies are mapped to prelinks and
  links in a deterministic way, but the weight of the configuration
  must be carried along, and observables are weighted averages. In the
  prelink temporal gauge, we first compute $Q_0$ and $Q_1$ as above and
  then have
  \begin{equation}
    V_0(x) =
    \begin{cases}
        I ,  & x_0 < L_0, x_1 < L_1\\
        Q_0 ,  & x_0 < L_0, x_1 = L_1\\
    \end{cases} .
  \end{equation}
  From the definition of holonomy we obtain
  \begin{equation}
    V_1(x) =
    \begin{cases}
        h_{1,0}(x) ,  & x_0 < L_0, x_1 < L_1\\
        h_{1,0}(x) Q_0 ,  & x_0 < L_0, x_1 = L_1\\
        Q_1\ h_{1,0}(x - L_0 \hat{0}), & x_0=L_0,\ x_1 < L_1
    \end{cases} ,
  \end{equation}
  where the last row is obtained by extending $V_1$ to the area with
  $x_0 = L_0$. Once $V_1(x)$ is obtained, we use $V_0(x) = h_{10}^\dagger(x)
  V_1(x)$, which completes the map from holonomy variables to prelinks.
  Note that the values of $V_0$ and $V_1$ at $(L_0, L_1)$
  do not matter here, since no link of the physical lattice uses that corner.
  The links then follow from eq.~\eqref{eq:U-from-V}. If needed, one can
  also perform a full gauge transformation of all sites to eliminate the
  effects of gauge fixing.
\end{enumerate}
\section{Simulations}
\label{sec:simulations}

\begin{figure*}[t]
\centering
\begin{minipage}{0.5\linewidth}
  \centering
  \includegraphics[width=1\linewidth, trim=0cm 0cm 0cm 1cm, clip]{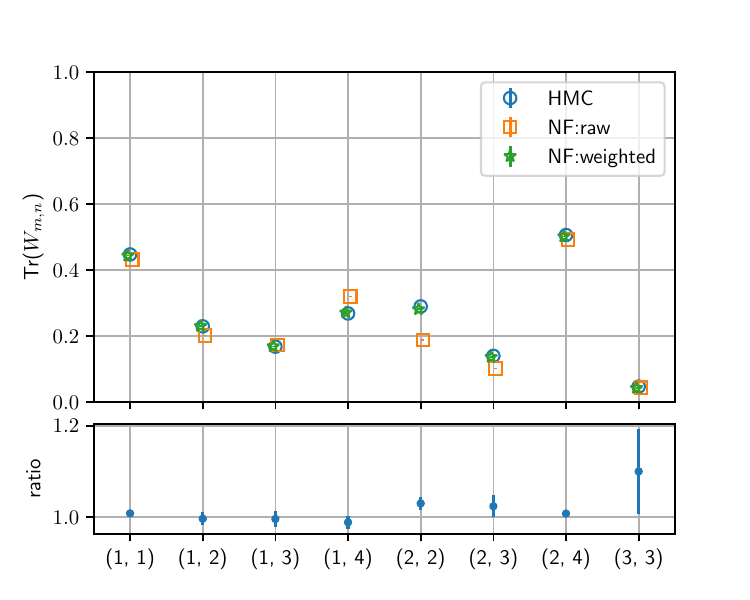}\par
  \small $2\times2$, $\beta=2$
\end{minipage}\hfill
\begin{minipage}{0.5\linewidth}
  \centering
  \includegraphics[width=1\linewidth, trim=0cm 0cm 0cm 1cm, clip]{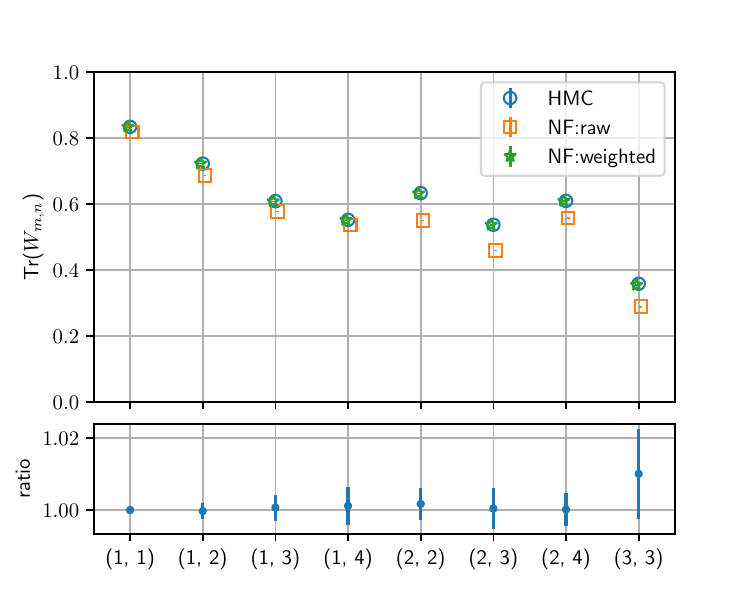}\par
  \small $2\times2$, $\beta=8$
\end{minipage}\\[6pt]
\begin{minipage}{0.5\linewidth}
  \centering
  \includegraphics[width=\linewidth]{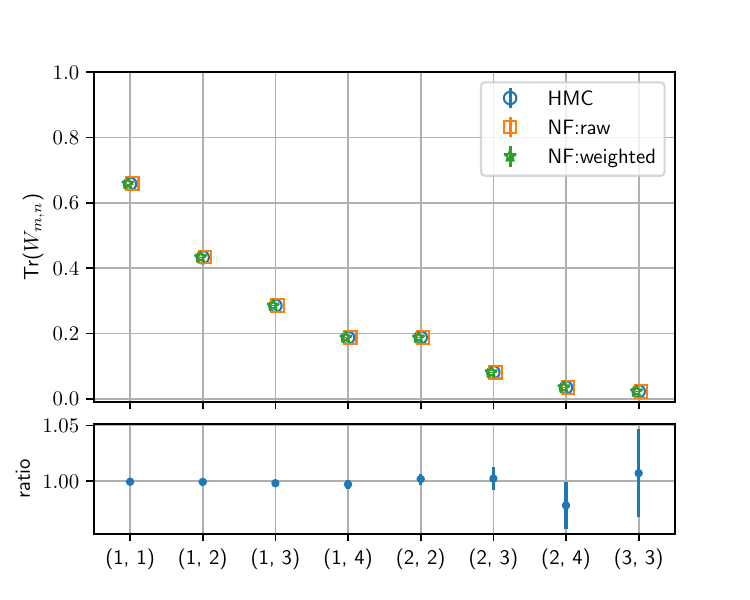}\par
  \small $32\times32$, $\beta=4$
\end{minipage}\hfill
\begin{minipage}{0.5\linewidth}
  \centering
  \includegraphics[width=\linewidth]{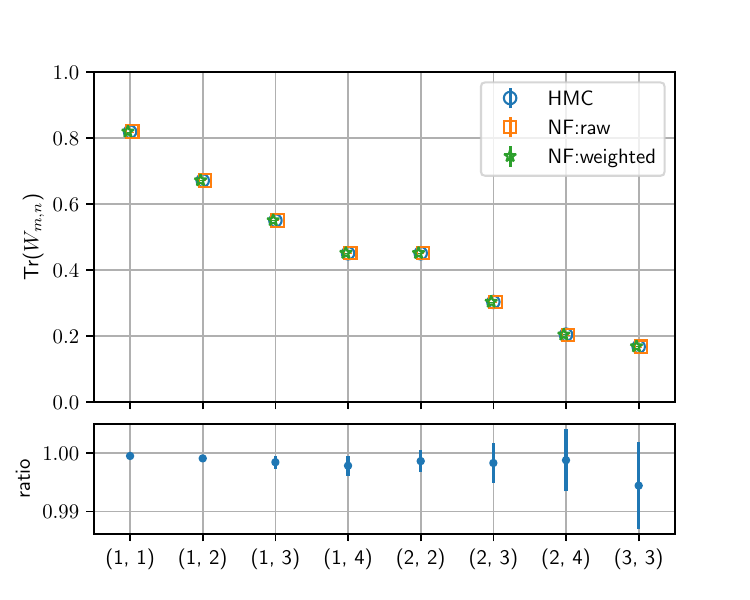}\par
  \small $32\times32$, $\beta=8$
\end{minipage}
\caption{$\SU(2)$: $\Tr(W_{m,n})$ from HMC, raw flow samples
(NF:raw), and reweighted flow samples (NF:weighted); ratio panels show
NF:weighted against HMC.}
\label{fig:wilson-su2}
\end{figure*}

\begin{figure*}[t]
\centering
\begin{minipage}{0.5\linewidth}
  \centering
  \includegraphics[width=1\linewidth, trim=0cm 0cm 0cm 1cm, clip]{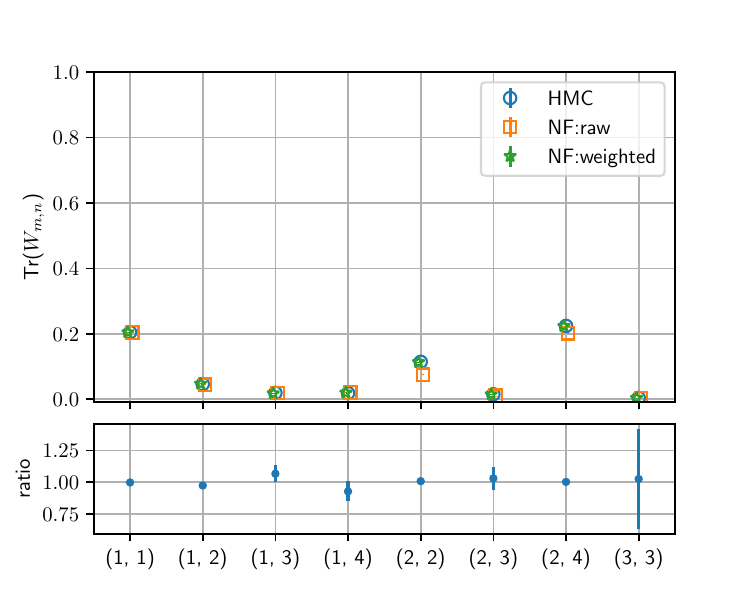}\par
  \small $2\times2$, $\beta=3$
\end{minipage}\hfill
\begin{minipage}{0.5\linewidth}
  \centering
  \includegraphics[width=1\linewidth, trim=0cm 0cm 0cm 1cm, clip]{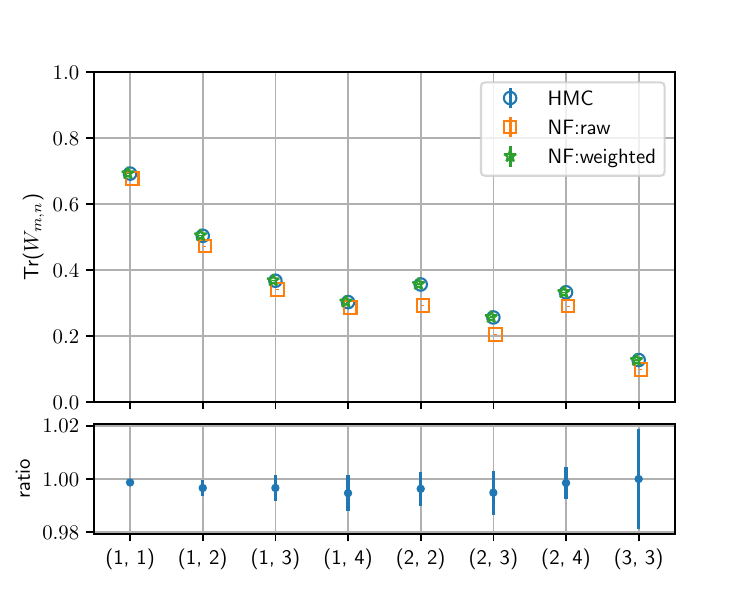}\par
  \small $2\times2$, $\beta=12$
\end{minipage}\\[6pt]
\begin{minipage}{0.5\linewidth}
  \centering
  \includegraphics[width=\linewidth]{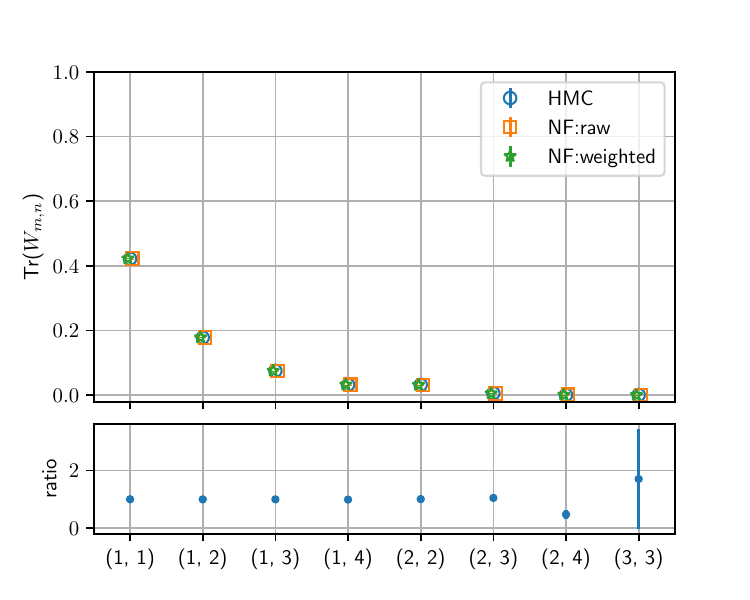}\par
  \small $32\times32$, $\beta=6$
\end{minipage}\hfill
\begin{minipage}{0.5\linewidth}
  \centering
  \includegraphics[width=\linewidth]{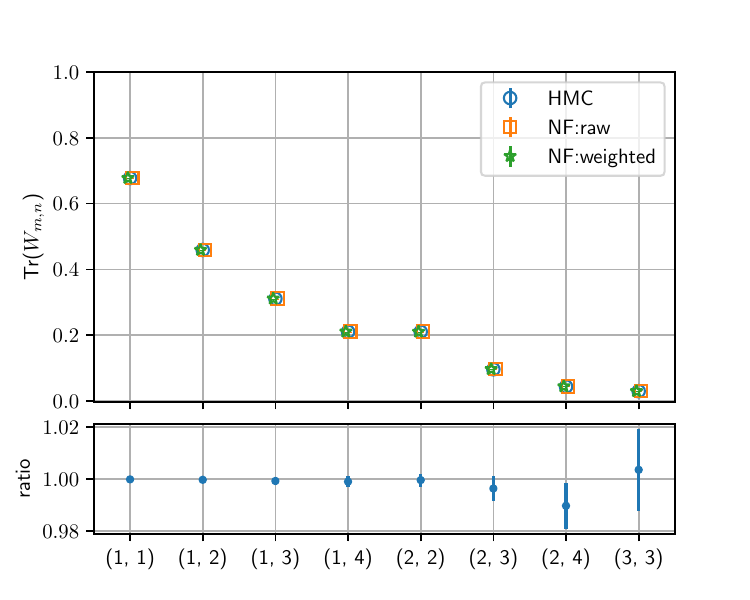}\par
  \small $32\times32$, $\beta=12$
\end{minipage}
\caption{$\SU(3)$: same comparison as fig.~\ref{fig:wilson-su2}.}
\label{fig:wilson-su3}
\end{figure*}

\begin{figure*}[t]
\centering
\includegraphics[width=1\linewidth]{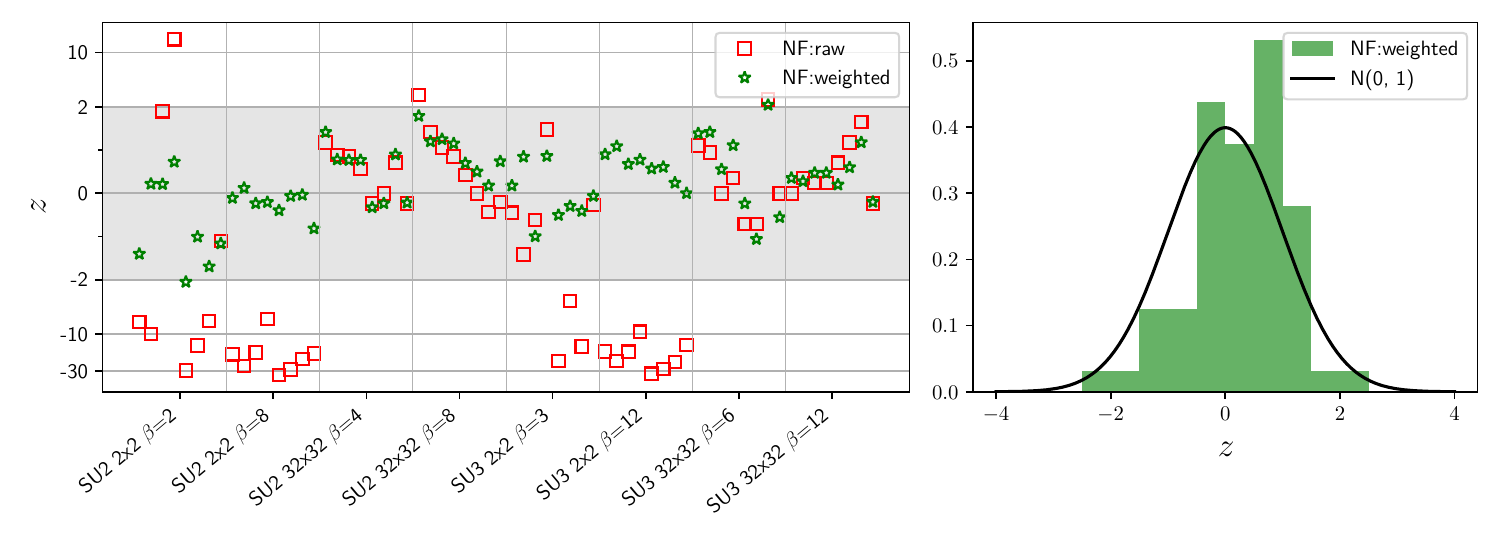}
\caption{Normalized deviation $z=(O_{\rm NF}-O_{\rm HMC})/\sqrt{\sigma_{\rm NF}^2+\sigma_{\rm HMC}^2}$
of the Wilson loops $\Tr(W_{m,n})$ of figs.~\ref{fig:wilson-su2}
and~\ref{fig:wilson-su3} (eight loops per setting, in the order $(1,1)$,
$(1,2)$, $(1,3)$, $(1,4)$, $(2,2)$, $(2,3)$, $(2,4)$, $(3,3)$), for NF:raw and
NF:weighted. The vertical axis is linear for $|z|<2$ (shaded) and
logarithmic beyond. Right: histogram of the $64$ weighted deviations against a
unit normal.}
\label{fig:pulls}
\end{figure*}

The single-plaquette distribution is a class function, so the normalizing flow acts only
on the conjugacy-class angles of the plaquette; one angle for $\SU(2)$ and two
for $\SU(3)$.
For the transformation, we employ rational-quadratic splines (20 segments for the
first angle and, for $\SU(3)$, 10 segments with a symmetry constraint for the
second).
The flow is trained in a self-learning manner, i.e., from its own samples with no external data, to reproduce the single-plaquette distribution at the given coupling. Training has two stages of $2000$ epochs with batch sizes $64$ and $1024$.
The conditional sampler for the $\SU(3)$ corner is described above in
sec.~\ref{sec:commutator-sampling-su3}.

Each normalizing flow result uses $n_\text{cfg}=16384$ configurations for the $2\times2$ lattices and
$n_\text{cfg}=1024$ for the $32\times32$ lattices.
The HMC results come from the same number of independent chains, each started
from a hot configuration and evolved for $100$ trajectories,
keeping only the final configuration of each chain, so that the HMC
configurations are statistically independent.
Each trajectory is a leapfrog integration of length $1$ with $20$ steps
(step size $0.05$), followed by a Metropolis accept/reject
step; the same step size is used for all settings and is not tuned.
The final acceptance rates range from $0.998$ ($\SU(2)$, $2\times2$, $\beta=2$)
to $0.72$ ($\SU(3)$, $32\times32$, $\beta=12$).

We validate the construction of the above section
for $\SU(2)$ and $\SU(3)$ gauge groups on $2\times2$ and $32\times32$ lattices, at one small and one
large coupling $\beta$ each.
For validation we check the results directly
against HMC.
Figures~\ref{fig:wilson-su2} and~\ref{fig:wilson-su3} compare the trace of Wilson loops
for a range of loop sizes and for the eight lattice settings.
We use ``NF:raw'' and ``NF:weighted'' to denote unweighted and weighted normalizing-flow results, respectively.
The figures also show the ratio of the weighted results against the HMC results.

To quantify the agreement we use the normalized deviation
\begin{equation}
    z = (O_{\rm NF}-O_{\rm HMC})/\sqrt{\sigma_{\rm NF}^2+\sigma_{\rm HMC}^2}
\end{equation}
of each of the $64$ loops $\Tr(W_{m,n})$ in the figures.
As illustrated in  fig.~\ref{fig:pulls},
for the $2\times2$ lattices the unweighted values differ from HMC by many
standard errors (root-mean-square of $z$ up to $22$, maximum of $|z|$ up to $34$), whereas the weighted
values agree well with it: over all $64$ loops the root-mean-square of $z$ is $0.86$ and
$|z|\lesssim2.1$. On the $32\times32$ lattices the raw results are already
largely compatible with HMC at the statistics used ($n_\text{cfg}=1024$).
The loops of a setting are computed from the same configurations and are
correlated, so these numbers are indicative rather than a formal test.

Table~\ref{tab:plaquette} gives the
plaquette expectation value $\Tr(W_{1,1})$ for all eight cases.
The effective sample size (ESS) of $n_\text{cfg}$ configurations with weights $w_i$ is
\begin{equation}
  \frac{\bigl(\sum_i w_i\bigr)^2}{n_\text{cfg}\sum_i w_i^2}\,,
\end{equation}
which equals $1$ for uniform weights.
The ESS values associated with the weights fluctuate from one run to another, and we
simply report a single run for each case in the table.

\begin{table}[t]
\centering
\begin{tabular}{ll c cccc}
\hline\hline
Group & Size & $\beta$ & HMC & NF:raw & NF:weighted & ESS \\
\hline
$\SU(2)$ & $2\times2$   & 2  & 0.4471(16) & 0.4312(16) & 0.4434(21) & 0.627 \\
$\SU(2)$ & $2\times2$   & 8  & 0.8340(6)  & 0.8186(6)  & 0.8339(7)  & 0.600 \\
$\SU(2)$ & $32\times32$ & 4  & 0.6579(3)  & 0.6584(3)  & 0.6585(3)  & 0.684 \\
$\SU(2)$ & $32\times32$ & 8  & 0.8191(1)  & 0.8195(1)  & 0.8195(2)  & 0.564 \\
$\SU(3)$ & $2\times2$   & 3  & 0.2046(11) & 0.2039(11) & 0.2049(13) & 0.724 \\
$\SU(3)$ & $2\times2$   & 12 & 0.6921(6)  & 0.6779(6)  & 0.6930(8)  & 0.639 \\
$\SU(3)$ & $32\times32$ & 6  & 0.4222(2)  & 0.4226(3)  & 0.4227(3)  & 0.673 \\
$\SU(3)$ & $32\times32$ & 12 & 0.6777(2)  & 0.6777(2)  & 0.6778(2)  & 0.747 \\
\hline\hline
\end{tabular}
\caption{Plaquette expectation value $\Tr(W_{1,1})$ for all eight
cases of figs.~\ref{fig:wilson-su2}--\ref{fig:wilson-su3}: HMC, raw
flow samples, and importance-reweighted flow samples. Digits in
parentheses are the error on the last digits. The last column is the ESS of the final weights, i.e., the weight $\rho(Z(h))$
times the conditional weight of the corner sampler, which is unity for
$\SU(2)$ (secs.~\ref{sec:commutator-sampling}, \ref{sec:workflow}). The ESS values
fluctuate from one run to another, especially for the $32\times32$ lattices,
and we report a single run for each case.}
\label{tab:plaquette}
\label{tab:ess}
\end{table}
\section{Summary and concluding remarks}
\label{sec:conclusion}

We have sampled two-dimensional $\SU(N)$ lattice gauge theory with the
Wilson plaquette action from independently sampled plaquettes, using the
prelink and holonomy parametrizations.
The holonomy parametrization maps the sampled plaquette values to link variables,
at the cost of a weight that must be carried by the resulting configurations.
The key ingredient is the boundary condition.
With periodic prelinks the holonomies must satisfy a consistency
condition at the four corners of the extended lattice, whereas with twisted
boundary conditions on the prelinks this condition becomes a conditional
sampling problem for $(X,Y)$ given the group commutator
$Z=XYX^\dagger Y^\dagger$, at the cost of the weight $\rho(Z)$. In our tests
for $\SU(2)$ and $\SU(3)$ on $2\times2$ and $32\times32$ lattices, the
effective sample size of the final weights remains moderate; in the runs reported
here it lies between $0.56$ and $0.75$, with no clear degradation from $2\times2$
to $32\times32$.
We restrict to the Wilson plaquette action, for which the
action in the holonomy variables is a sum of independent plaquettes; for more
general actions the holonomy formulation need not be of the open-boundary form.

To solve the corner problem we introduced an encoder,
$(X,Y)\to(D,\Omega,Z)$, together with its decoder for $\SU(2)$ and $\SU(3)$.
In these coordinates the measure factorizes into the marginal density $\rho(Z)$ of $Z$ and the conditional law of $(D,\Omega)$ given $Z$.
The distribution of $D$ is uniform and independent of $Z$, and, by conjugation invariance, the remaining dependence on $Z$ is only through its $N-1$ independent
eigen-angles.

For $\SU(2)$ everything is available in closed form: the marginal density of
$Z$, $\rho(\theta)=(\pi-\theta)/(2\sin\theta)$, and the conditional law of
$(D,\Omega)$ given $Z$, whose only nontrivial part has an explicit cumulative
distribution function that can be inverted. The corner sampler is therefore
exact, and the effective sample size is determined by the density of $Z$ alone.
This density depends on the lattice size and the coupling, but $Z$ approaches
the Haar distribution for asymptotically large lattices, where the effective
sample size approaches $6/\pi^2\approx0.61$.

For $\SU(3)$ the marginal density of $Z$ is also available in closed form,
eq.~\eqref{eq:commutator-density}. The conditional law of $(D,\Omega)$ given $Z$, however, is not yet: our
sampler is a trained approximation, whose weight is computed exactly and whose
effective sample size at fixed $Z$ is about $0.98$. We are still working on
constructing an exact conditional sampler for $\SU(3)$.

The closed form of the density of $Z$, eq.~\eqref{eq:commutator-density}, was
derived above for $N=3$ by summing the character series. Up to the factor
$1/N$, it is the ratio of the Vandermonde determinants of the eigen-angles and
of the eigenvalues. Interestingly, it also holds for $N=2$. Whether a closed
form of this type exists for $\SU(N)$ with $N > 3$ is left for future work.

The holonomy parametrization can also be used in higher dimensions, but there
the boundary holonomies are subject to many additional constraints, and
resolving them is not straightforward. We leave this for future work; one
option is to project the HMC force onto the constraint
surface.
Another advantage of the holonomy parametrization is that it can be
used to construct gauge-equivariant networks for both flow-based and
diffusion models.

\section*{Note added}
While this paper was being finalized, ref.~\cite{Singha:2026qyx}
appeared, which explores related ideas for $\mathrm U(1)$ and $\SU(2)$ gauge
theories: it samples directly in plaquette space with a multilevel
normalizing-flow construction that satisfies the Bianchi identities
exactly. It treats $\mathrm U(1)$ in two and four dimensions and
$\SU(2)$ in two dimensions. It works on a periodic lattice and, to the best of our
knowledge, does not use twisted boundary conditions.

\section*{Code availability}
The notebooks used to produce the results of this paper, including the trained
normalizing-flow weights, are available at
\url{https://github.com/jkomijani/holonomy-2d-gauge-notebooks}. They use the
packages \texttt{lattice\_ml}, version 0.2.0
(\url{https://github.com/jkomijani/lattice_ml}), and \texttt{normflow}, version
3.0.0 (\url{https://github.com/jkomijani/normflow}).

\section*{Use of AI tools}
Claude (Anthropic) and ChatGPT (OpenAI) were used to improve the text.
Claude was also used for implementing the Euler-angle parametrization of $\SU(3)$,
which the author checked for correctness. Claude found a closed-form expression for the
marginal distribution of the group commutator $Z$ for $N=3$.
That computation, however, was overwhelming; starting from the closed form
found by Claude, the author reverse-engineered a simpler derivation
(sec.~\ref{sec:commutator-density}).
The author is responsible for the content of the paper.

\begin{acknowledgments}
The author thanks Benjamin Izett for introducing the string formulation of
refs.~\cite{Bars:1978ry,Batrouni:1984rb} and for helpful discussions about it.
\end{acknowledgments}

\bibliographystyle{apsrev4-1}
\bibliography{references.bib}

@article{Wilson:1974sk,
    author = "Wilson, Kenneth G.",
    title = "{Confinement of Quarks}",
    reportNumber = "CLNS-262",
    doi = "10.1103/PhysRevD.10.2445",
    journal = "Phys. Rev. D",
    volume = "10",
    pages = "2445--2459",
    year = "1974"
}

@article{Migdal:1975zg,
    author = "Migdal, Alexander A.",
    title = "{Recursion equations in gauge field theories}",
    reportNumber = "PRINT-75-1043 (LANDAU-INST)",
    journal = "Sov. Phys. JETP",
    volume = "42",
    pages = "413--418",
    year = "1975"
}

@article{Drouffe:1983fv,
    author = "Drouffe, Jean-Michel and Zuber, Jean-Bernard",
    title = "{Strong Coupling and Mean Field Methods in Lattice Gauge Theories}",
    reportNumber = "PRINT-83-0659 (CERN)",
    doi = "10.1016/0370-1573(83)90034-0",
    journal = "Phys. Rept.",
    volume = "102",
    pages = "1",
    year = "1983"
}

@article{Rusakov:1990rs,
    author = "Rusakov, B. E.",
    title = "{Loop averages and partition functions in U(N) gauge theory on two-dimensional manifolds}",
    doi = "10.1142/S0217732390000780",
    journal = "Mod. Phys. Lett. A",
    volume = "5",
    pages = "693--703",
    year = "1990"
}

@article{Witten:1991we,
    author = "Witten, Edward",
    title = "{On quantum gauge theories in two-dimensions}",
    doi = "10.1007/BF02100009",
    journal = "Commun. Math. Phys.",
    volume = "141",
    pages = "153--209",
    year = "1991"
}

@article{Duane:1987de,
    author = "Duane, S. and Kennedy, A. D. and Pendleton, B. J. and Roweth, D.",
    title = "{Hybrid Monte Carlo}",
    doi = "10.1016/0370-2693(87)91197-X",
    journal = "Phys. Lett. B",
    volume = "195",
    pages = "216--222",
    year = "1987"
}

@article{Luscher:2009eq,
    author = "L{\"u}scher, Martin",
    title = "{Trivializing maps, the Wilson flow and the HMC algorithm}",
    eprint = "0907.5491",
    archivePrefix = "arXiv",
    primaryClass = "hep-lat",
    reportNumber = "CERN-PH-TH-2009-118",
    doi = "10.1007/s00220-009-0953-7",
    journal = "Commun. Math. Phys.",
    volume = "293",
    pages = "899--919",
    year = "2010"
}

@article{Kanwar:2020xzo,
    author = "Kanwar, Gurtej and Albergo, Michael S. and Boyda, Denis and Cranmer, Kyle and Hackett, Daniel C. and Racani\`ere, S\'ebastien and Rezende, Danilo Jimenez and Shanahan, Phiala E.",
    title = "{Equivariant flow-based sampling for lattice gauge theory}",
    eprint = "2003.06413",
    archivePrefix = "arXiv",
    primaryClass = "hep-lat",
    reportNumber = "MIT-CTP/5181",
    doi = "10.1103/PhysRevLett.125.121601",
    journal = "Phys. Rev. Lett.",
    volume = "125",
    number = "12",
    pages = "121601",
    year = "2020"
}

@article{Boyda:2020hsi,
    author = "Boyda, Denis and Kanwar, Gurtej and Racani\`ere, S\'ebastien and Rezende, Danilo Jimenez and Albergo, Michael S. and Cranmer, Kyle and Hackett, Daniel C. and Shanahan, Phiala E.",
    title = "{Sampling using $SU(N)$ gauge equivariant flows}",
    eprint = "2008.05456",
    archivePrefix = "arXiv",
    primaryClass = "hep-lat",
    reportNumber = "MIT-CTP/5228",
    doi = "10.1103/PhysRevD.103.074504",
    journal = "Phys. Rev. D",
    volume = "103",
    number = "7",
    pages = "074504",
    year = "2021"
}

@article{Favoni:2020reg,
    author = "Favoni, Matteo and Ipp, Andreas and M{\"u}ller, David I. and Schuh, Daniel",
    title = "{Lattice Gauge Equivariant Convolutional Neural Networks}",
    eprint = "2012.12901",
    archivePrefix = "arXiv",
    primaryClass = "hep-lat",
    doi = "10.1103/PhysRevLett.128.032003",
    journal = "Phys. Rev. Lett.",
    volume = "128",
    number = "3",
    pages = "032003",
    year = "2022"
}

@article{Bacchio:2022vje,
    author = "Bacchio, Simone and Kessel, Pan and Schaefer, Stefan and Vaitl, Lorenz",
    title = "{Learning trivializing gradient flows for lattice gauge theories}",
    eprint = "2212.08469",
    archivePrefix = "arXiv",
    primaryClass = "hep-lat",
    doi = "10.1103/PhysRevD.107.L051504",
    journal = "Phys. Rev. D",
    volume = "107",
    number = "5",
    pages = "L051504",
    year = "2023"
}

@article{Aarts:2026zzr,
    author = {Aarts, Gert and Habibi, Diaa E. and Ipp, Andreas and M{\"u}ller, David I. and Ranner, Thomas R. and Wang, Lingxiao and Wang, Wei and Zhu, Qianteng},
    title = "{Generalizable Equivariant Diffusion Models for Non-Abelian Lattice Gauge Theory}",
    eprint = "2601.19552",
    archivePrefix = "arXiv",
    primaryClass = "hep-lat",
    reportNumber = "RIKEN-iTHEMS-Report-26",
    month = "1",
    year = "2026"
}

@article{Alharazin:2026lcb,
    author = "Alharazin, H. and Panteleeva, J. Yu. and Sun, B. -D.",
    title = "{Diffusion models for SU(2) lattice gauge theory in two dimensions}",
    eprint = "2602.09045",
    archivePrefix = "arXiv",
    primaryClass = "hep-lat",
    doi = "10.1103/n3bh-gs98",
    journal = "Phys. Rev. D",
    volume = "114",
    number = "1",
    pages = "014522",
    year = "2026"
}

@article{Komijani:2025yjz,
    author = "Komijani, Javad and Marinkovic, Marina K.",
    title = "{Normalizing flows for SU(N) gauge theories employing singular value decomposition}",
    eprint = "2501.18288",
    archivePrefix = "arXiv",
    primaryClass = "hep-lat",
    doi = "10.22323/1.466.0050",
    journal = "PoS",
    volume = "LATTICE2024",
    pages = "050",
    year = "2025"
}

@article{Komijani:2026lan,
    author = "Komijani, Javad and Marinkovic, Marina K. and Turgut, Lara",
    title = "{Diffusion model for SU(N) gauge theories}",
    eprint = "2605.06134",
    archivePrefix = "arXiv",
    primaryClass = "hep-lat",
    month = "5",
    year = "2026"
}

@article{Komijani:2026mlx,
    author = "Komijani, Javad",
    title = "{Noise scheduling and linear dynamics in diffusion models on Lie groups}",
    eprint = "2605.17326",
    archivePrefix = "arXiv",
    primaryClass = "hep-lat",
    month = "5",
    year = "2026"
}

@article{Bars:1978ry,
    author = "Bars, Itzhak",
    title = "{A Map Between Corner and Link Operators in Lattice Gauge Theories}",
    reportNumber = "COO-3075-195",
    doi = "10.1016/0550-3213(79)90155-X",
    journal = "Nucl. Phys. B",
    volume = "149",
    pages = "39--48",
    year = "1979"
}

@article{Batrouni:1984rb,
    author = "Batrouni, Ghassan G. and Halpern, M. B.",
    title = "{String, Corner and Plaquette Formulation of Finite Lattice Gauge Theory}",
    reportNumber = "CLNS-84/602",
    doi = "10.1103/PhysRevD.30.1782",
    journal = "Phys. Rev. D",
    volume = "30",
    pages = "1782",
    year = "1984"
}

@article{Singha:2026qyx,
    author = "Singha, Ankur and Finkenrath, Jacob and Jansen, Karl and Arora, Vipul and Nakajima, Shinichi",
    title = "{Multilevel Plaquette-Space Sampling for Lattice Gauge Theories with Local Constraint Solves}",
    eprint = "2609.35722",
    archivePrefix = "arXiv",
    primaryClass = "hep-lat",
    month = "9",
    year = "2026"
}

@article{vaitl:2022path,
    author = "Lorenz Vaitl and Kim A. Nicoli and Shinichi Nakajima and Pan Kessel",
    title = "{Gradients should stay on Path: Better Estimators of the Reverse- and Forward KL Divergence for Normalizing Flows}",
    year = {2022},
    eprint = {2207.08219},
    archivePrefix = {arXiv},
    primaryClass = {cs.LG}
}

@article{Baaquie:1988su3,
    author = "Baaquie, Belal E.",
    title = "{Character functions of SU(3)}",
    doi = "10.1088/0305-4470/21/11/022",
    journal = "J. Phys. A",
    volume = "21",
    pages = "2651--2656",
    year = "1988"
}

@book{Greiner:1994sym,
    author = "Greiner, Walter and M{\"u}ller, Berndt",
    title = "{Quantum Mechanics: Symmetries}",
    edition = "2nd",
    publisher = "Springer",
    address = "Berlin, Heidelberg",
    isbn = "3-540-58080-8",
    year = "1994"
}

@article{Mulase:2002vol,
    author = "Mulase, Motohico and Penkava, Michael",
    title = "{Volume of representation varieties}",
    eprint = "math/0212012",
    archivePrefix = "arXiv",
    primaryClass = "math.QA",
    year = "2002"
}

@article{Griguolo:2022vdx,
    author = "Griguolo, Luca and Panerai, Rodolfo and Papalini, Jacopo and Seminara, Domenico",
    title = "{The phase diagram of $ T\overline{T} $-deformed Yang-Mills theory on the sphere}",
    eprint = "2209.06222",
    archivePrefix = "arXiv",
    primaryClass = "hep-th",
    reportNumber = "UUITP-38/22",
    doi = "10.1007/JHEP11(2022)078",
    journal = "JHEP",
    volume = "11",
    pages = "078",
    year = "2022"
}

\end{document}